\documentclass[prl,twocolumn,floatfix,a4paper,superscriptaddress,%
        nobalancelastpage]{revtex4}

\usepackage{amsmath,amsfonts,amssymb,graphics,graphicx,epsfig,color,bbm}
\usepackage{times}

\DeclareMathOperator{\spanned}{span}

\newcommand{\figgg}[2]{\ensuremath{\vcenter{\hbox{{\includegraphics[ scale=#1]{Fig/#2}}}}}}
\newcommand{\figgs}[2]{\ensuremath{\vcenter{\hbox{{\includegraphics[ scale=#1]{Fig/Fig2/#2}}}}}}

\newcommand{\R}{\mathbbm{R}}
\newcommand{\C}{\mathbb{C}}
\newcommand{\I}{\mathbb{I}}
\newcommand{\be}{\begin{equation}}
\newcommand{\ee}{\end{equation}}
\newcommand{\bea}{\begin{eqnarray}}
\newcommand{\eea}{\end{eqnarray}}
\newcommand{\ket}[1]{|#1\rangle}
\newcommand{\bra}[1]{\langle#1|}
\newcommand{\braket}[2]{\langle #1|#2\rangle}

\begin{document}


\newtheorem{theorem}{Theorem}
\newtheorem{proposition}{Proposition}

\newtheorem{lemma}{Lemma}

\newtheorem{definition}{Definition}
\newtheorem{corollary}{Corollary}

\newcommand{\proof}[1]{{\bf Proof.} #1 $\proofend$}

\title{Gapless Hamiltonians for the toric code using the PEPS formalism}

\author{Carlos Fern\'andez-Gonz\'alez}
\affiliation{Departamento de F\'{\i}sica de los Materiales,
Universidad Nacional de Educaci\'on a Distancia (UNED), 28040 Madrid,
Spain}
\affiliation{Departamento de An\'alisis Matem\'atico \& IMI, Universidad
Complutense de Madrid, 28040 Madrid, Spain}

\author{Norbert Schuch}
\affiliation{Institute for Quantum Information,
California Institute of Technology,
MC 305-16, Pasadena CA 91125, U.S.A.}

\author{Michael M. Wolf}
\affiliation{Department of Mathematics, Technische Universit\"at
M\"unchen, 85748 Garching, Germany}

\author{J. Ignacio Cirac}
\affiliation{Max-Planck-Institut f\"ur Quantenoptik, Hans-Kopfermann-Str.\
1, 85748 Garching, Germany} 

\author{David P\'erez-Garc\'{\i}a}
\affiliation{Departamento de An\'alisis Matem\'atico \& IMI, Universidad
Complutense de Madrid, 28040 Madrid, Spain}


\begin{abstract} 
We study Hamiltonians which have Kitaev's toric code as a ground state, and
show how to construct a Hamiltonian which shares the ground space of the
toric code, but which has gapless excitations with a continuous spectrum
in the thermodynamic limit.  Our construction is based on the framework of
Projected Entangled Pair States (PEPS), and can be applied to a large
class of two-dimensional systems to obtain gapless ``uncle Hamiltonians''.
\end{abstract}

\maketitle



\emph{Introduction.---}%
Since its introduction by Wen in the 80's, topological order has become a
central subject of research both in the condensed matter and quantum
information communities. The toric code, a many-body spin state originally
introduced by Kitaev in the context of topological quantum
computing~\cite{kitaev:toriccode}, represents a paradigmatic example of a
state with topological order. It is the ground state of a local,
frustration free Hamiltonian $H_\mathrm{TC}$ defined on a two-dimensional
lattice,  whose degeneracy depends on the topology of the space on which
it is defined.  This Hamiltonian is gapped, and it exhibits (abelian)
anyonic excitations.  The toric code also possesses long-range
entanglement (i.e., it cannot be created by local unitary operations out
of a product state), and its entanglement entropy contains a universal
part which can serve as a signature of its topological properties.  All
these properties are robust against local
perturbations~\cite{bravyi:tqo-long,bravyi:local-tqo-simple}.  Apart from
that, it can be considered as an error correcting code with non-local
encoding but local syndroms, and might therefore be useful as a quantum
memory or for fault tolerant quantum computing.

The toric code can also be efficiently described in the language of tensor
networks.  As other states with topological order, it is a Projected
Entangled Pair State (PEPS) of very low bond dimension,
$D=2$~\cite{verstraete:mbc-peps,verstraete:comp-power-of-peps}. PEPS
generalize Matrix Product States
(MPS)~\cite{fannes:FCS,perez-garcia:mps-reps} to spatial dimensions higher
than one, obey the area law for the entanglement entropy, and are believed
to efficiently represent the ground states of local spin and fermionic
Hamiltonians in lattices~\cite{hastings:mps-entropy-ent,kraus:fPEPS}.
Conversely, for any PEPS one can construct a frustration free \emph{parent
Hamiltonian} for which it is the ground
state~\cite{verstraete:comp-power-of-peps}, which allows us to relate a
given exotic quantum many-body state to physical Hamiltonians.  In fact,
$H_\mathrm{TC}$ is exactly such a parent Hamiltonian for the toric code,
and using this construction in the PEPS formalism, one can readily uncover
some of its most distinct properties~\cite{schuch:peps-sym}.  In the
same way, one can build parent Hamiltonians for many other strongly
correlated states, such as string-net models~\cite{wen:stringnets}, the
AKLT state~\cite{aklt}, resonating valence bond states, and others.  In
most of these cases, the resulting Hamiltonians are gapped above the
ground state space, which makes them robust against local
perturbations~\cite{michalakis:local-tqo-ffree}.

In this paper, we introduce an alternative way to construct Hamiltonians
corresponding to MPS and PEPS, which we term \emph{uncle Hamiltonians}. The uncle Hamiltonian
differs significantly from the parent Hamiltonian. While both Hamiltonians
share the same ground state subspace by construction, their spectra are
extremely different: As we prove, the uncle Hamiltonian is gapless and has
a continuous spectrum in the thermodynamic limit, which is in sharp
contrast to the gapped parent Hamiltonian. Our construction exploits the fact that the link between tensor networks and their associated parent Hamiltonians is not robust under generic perturbations \cite{chen:parentH-robustness} for a large class of interesting MPS and PEPS, in particular for systems with symmetry breaking and topological order.

Our findings are interesting from several perspectives. First, they show
that the association between PEPS and Hamiltonians is more ambiguous than
generally believed.  Second, it illustrates that care must be taken when trying to define topological
order in terms of properties of the ground
state alone, such as its topological
entropy~\cite{levin:topological-entropy,kitaev:topological-entropy}, as
the same quantum state can appear as a ground state of both a gapped
(topological) and a gapless (unstable) Hamiltonian. Finally, it also
provides a clear example of a gapless system which nevertheless does not
exhibit any critical (or even finite-range) correlations.

\emph{Uncle Hamiltonian for the GHZ state.---}%
We start by explaining our construction for the GHZ state in order to
introduce the key concepts.

A state $\ket{\psi}\in (\C ^d)^{\otimes L}$ is called a (translationally
invariant) Matrix Product State (MPS) if it can be written as $$
\ket{M(A)} = \sum _{i_1,...,i_L} \mathrm{tr}[A_{i_1} \cdot \cdot \cdot
A_{i_L}] \ket{ i_1,...,i_L},$$
where the $A_i$ are $D\times D$ matrices, $D$ being called the
\emph{bond dimension}. These matrices can be thought of as a tensor $A$ with three
indices $(A_i)_{\alpha\beta}$, two of them ($\alpha$, $\beta$) being the
matrix indices (``virtual indices'') and the third index ($i$)
corresponding to the physical spin (``physical index'').

The unnormalized GHZ state on $n$ particles can be expressed as an MPS as
follows:
\[
\ket{\mathrm{GHZ}}= \!\!\sum _{i_1,...,i_n} \!\!
	\mathrm{tr}[A_{i_1} \cdots A_{i_n}] \ket{ i_1...i_n}=
    \ket{00\dots0}+\ket{11\dots1},
\]
where  $i_j \in \{0,1\}$ and $A_0=
\left( \begin{smallmatrix}
1 & 0 \\
0 & 0
\end{smallmatrix} \right)$, $A_1=\left( \begin{smallmatrix}
0 & 0 \\
0 & 1
\end{smallmatrix} \right)$.

A parent Hamiltonian $H=\sum_i h_\mathrm{loc}$ of an MPS is
obtained as a sum of local orthogonal projections
$h_{\mathrm{loc}}=h_{i-1,i,i+1}$ acting on three consecutive
sites, each of them with
kernel~\cite{perez-garcia:mps-reps,schuch:peps-sym}.
\[
\spanned\Big\{\sum_{i_1i_2i_3} \bra{i} A_{i_1} A_{i_2} A_{i_3} \ket{j}
\ket{i_1i_2i_3},\ i,j\in \{0,1\}\Big\}\ ;
\]
for the GHZ state, $\mathrm{ker}\,
h_{\mathrm{loc}}=\spanned\{\ket{000},\ket{111}\}$ \footnote{The parent Hamiltonian is usually obtained by considering local projectors on just two sites, but we need three sites for our construction in this specific case. A discussion on the need of either two or three sites and its relationship with bond and physical dimensions can be found in [17].}

The parent Hamiltonian $H$ is frustration free since its ground space is
the intersection of these kernels. The GHZ state lies in the ground space,
which is 2-dimensional and is spanned by the states $\ket{0}^{\otimes n}$
and $\ket{1}^{\otimes n}$, and the Hamiltonian has an spectral gap between
the ground space and the rest of energy levels.

Let us now perturb the GHZ state in the MPS representation, by considering
small random linear perturbations of the matrices $A_i$, 
\[
A_{0}^\varepsilon =
A_0 +\varepsilon \left( \begin{array}{cc}
a_0 & b_0 \\
c_0 & d_0
\end{array} \right), \ A_1^{\varepsilon} = A_1+\varepsilon \left( \begin{array}{cc}
a_1 & b_1 \\
c_1 & d_1
\end{array} \right)\ .
\]

The parent Hamiltonian $H^\varepsilon$ corresponding to this new MPS is
the sum of a new local projector $h^\varepsilon_{\mathrm{loc}}$ with 
\[
\ker h^\varepsilon_{\mathrm{loc}}=\spanned\Big\{\sum_{i_1i_2i_3}
\bra{i} A^\varepsilon_{i_1} A^\varepsilon_{i_2} A^\varepsilon_{i_3}
\ket{j} \ket{i_1i_2i_3},\ i,j\in \{0,1\}\Big\}\ .
\]
This kernel is spanned by the vectors 
\begin{eqnarray*}
&\ket{000}+O(\varepsilon)\ ,\\
&\ket{111}+O(\varepsilon)\ , \\ 
&\varepsilon\big[b_0 \ket{000} +(b_0+b_1)  (\ket{001} + \ket{011})+
b_1\ket{111}\big]+O(\varepsilon^2)\ ,\\ 
&\varepsilon\big[c_0 \ket{000} +(c_0+c_1) ( \ket{100} +  \ket{110})+
c_1\ket{111}\big]+O(\varepsilon^2)\ ,
\end{eqnarray*}

\noindent or equivalently by the vectors 
\begin{align*}
\ket{000} + O(\varepsilon)\ ,\quad 
\ket{0\!+\!1}+ O(\varepsilon)\ ,\\
\ket{111}+O(\varepsilon)\ ,
\quad  \ket{1\!+\!0} + O(\varepsilon)\ ,
\end{align*}
as long as $b_0+b_1\ne 0$ and $c_0+c_1\ne 0$, which holds for almost every
perturbation. [$\ket{0\!+\!1}$ denotes $\ket0\ket+\ket1$, etc., where $\ket{+}=(\ket{0}+\ket{1})/\sqrt{2}$]

\begin{figure}[t]
  \includegraphics[width=6cm]{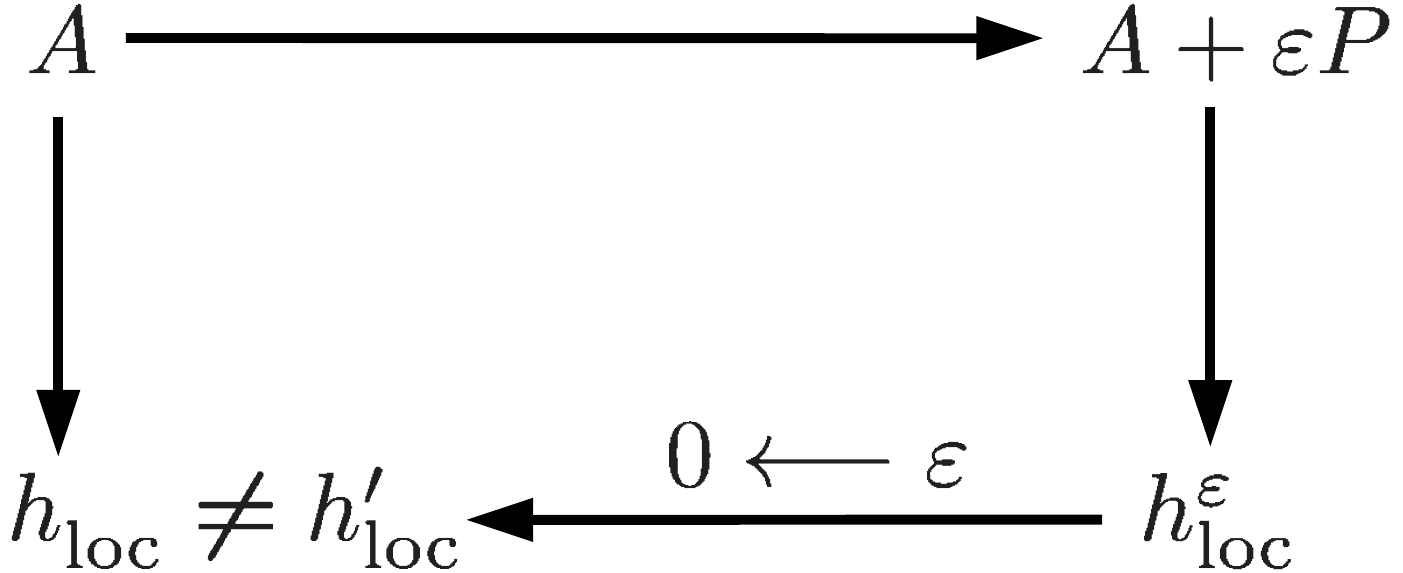}
\caption{ \label{fig1}
Construction of the uncle Hamiltonian. Any MPS and PEPS tensor $A$ induces
a corresponding parent Hamiltonian $h_\mathrm{loc}$. The \emph{uncle
Hamiltonian} is constructed by perturbing $A\rightarrow A+\varepsilon P$,
computing its parent Hamiltonian $h_\mathrm{loc}^\epsilon$, and finally
taking $\varepsilon$ to zero.  As we show, the resulting uncle Hamiltonian
$h_\mathrm{loc}'=\lim h_\mathrm{loc}^\varepsilon$ can be very different
from the parent Hamiltonian $h_\mathrm{loc}$.}
\end{figure}

As we let $\varepsilon$ tend to $0$, this local projector does not converge
to the original $h_{i-1,i,i+1}$. Instead, it converges to a projector with
kernel $\spanned\{\ket{000},\ket{0\!+\!1},\ket{1\!+\!0},\ket{111}\}$,
which we denote by $h'_{\mathrm{loc}}=h'_{i-1,i,i+1}$ for the corresponding
sites.  The resulting global Hamiltonian $H'=\sum_i h'_{i-1,i,i+1}$ is the
one we will call the \emph{uncle Hamiltonian}.  As
$\mathrm{ker}\,h_\mathrm{loc}\subset \mathrm{ker}\,h'_\mathrm{loc}$, $H'$
has all the ground states of $H$ and is thus frustration free.  On the
other hand, the presence of the vector $\ket{0\!+\!1}$ in the ground state
subspace also allows for zero-momentum superpositions of ``domain walls''
between domains of $0$'s and $1$'s,
$\dots+\ket{\dots001\dots}+\ket{\dots011\dots}+\cdots$, and correspondingly for
$\ket{1\!+\!0}$. However, it is easy to see that these configurations
cannot exist in the ground space given periodic boundary conditions (there
need to be an even number of domain walls, but they are not allowed to
meet), and thus, the ground state subspace for $H'$ is the same as for the
parent Hamiltonian $H$. On the other hand, $H'$ is gapless in
the thermodynamic limit, and moreover, its spectrum is the whole positive
real line $\R^+$. This can be
proven utilizing the ``domain wall superpositions'' mentioned above, by 
using the unnormalized states 
\be\label{gapvector} 
\ket{\varphi_{r,A}}=
\sum_{\substack{{1\le i<j\le A,}\\{\,2\le j-i \le
r}}} \ket{\dots 0} \ket{0^10\dots 0^i11\dots 1^j0\dots 0^A}\ket{0\dots},
\ee
where the superscripts indicate the position of the
corresponding site. The $\ket{\phi_{r,A}}$ are orthogonal to the ground space and have
energy as close to $C/(r-1)$ as desired if we allow both the chain length and $A$ grow, for some given constant $C$ and
any $r$, which implies the existence of low eigenvalues tending to 0. The
locality of the uncle Hamiltonian renders it possible to concatenate
approximate eigenvectors in the thermodynamic limit and therefore
to conclude that the sum of two elements in the spectrum is also in the
spectrum, which finally allows to prove that the spectrum of $H'$ becomes
continuous in the thermodynamic limit, $\sigma(H')=\R^+$.  Moreover, the
spectra of $H'$ acting on finite size chains tend to be dense on the
positive real line as the size of the chain grows.  A detailed
analysis is given in Ref.~\cite{fernandez:1d-uncle}, where it is also
shown that in one dimension, this behavior occurs for any MPS with
degenerate parent Hamiltonian (i.e., non-injective MPS). In contrast, for systems with unique ground states (i.e., injective MPS)
the parent Hamiltonian is robust under perturbations. However, one can
still construct gapless uncle Hamiltonians by taking non-injective MPS
representations of these states, but in this case the similarities between
parent and uncle Hamiltonians are weaker \cite{fernandez:1d-uncle}.

\begin{figure}[t]
  \includegraphics[width=4cm]{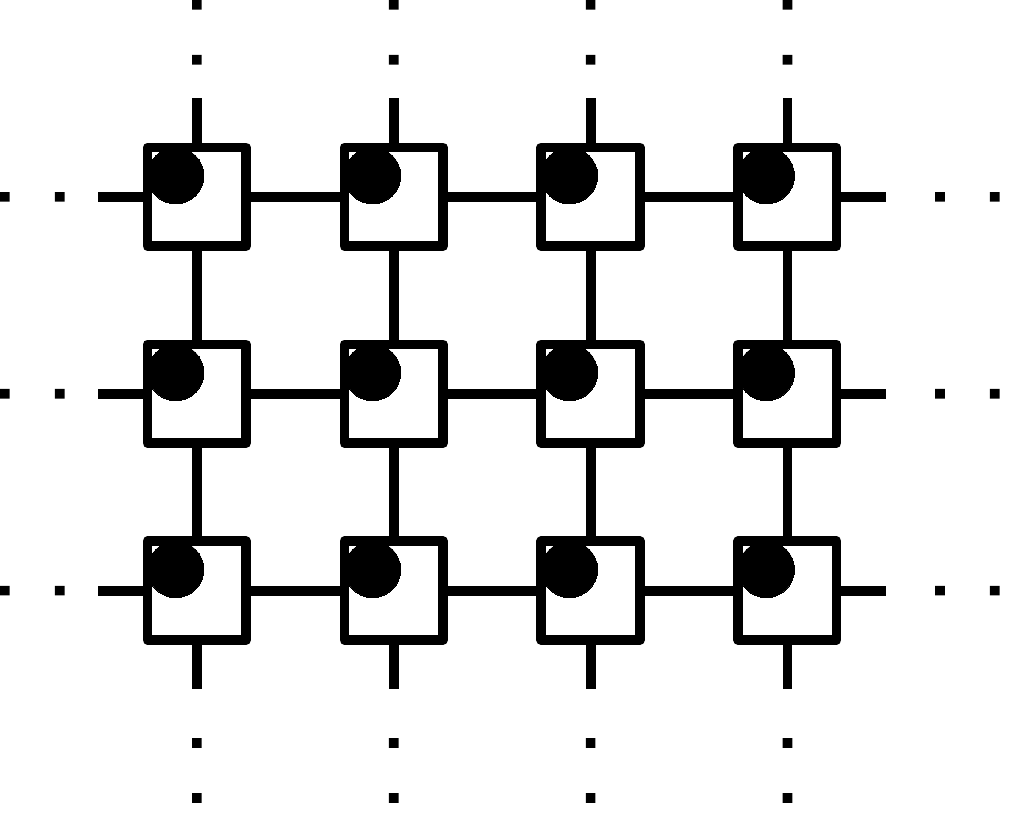}\\
  \caption{Graphical description of PEPS.\label{fig2}}
\end{figure}

\emph{The toric code as a PEPS.---}%
Projected Entangled Pair States (PEPS) are the natural generalization of
MPS to general lattices. For simplicity, we restrict to square lattices.
Then, the three-index tensors $A$ have to be replaced by five-index
tensors, with four virtual indices and one physical index. The virtual indices
of each tensor are contracted with the corresponding indices of the adjacent tensors as
depicted in Fig.~\ref{fig2}, where connected lines denote the contraction
of indices.  The physical index will be denoted by a black dot in the
upper left corner of each tensor, and should be thought of as a tensor leg
pointing out of the paper.

Under certain conditions on the tensors~\cite{schuch:peps-sym}, a parent
Hamiltonian can be constructed by considering local
projections $h_\mathrm{loc}$ for every $2\times 2$ region onto the
orthogonal complement of the space 
\be
\label{eq:parent2d}
\ker h_{\mathrm{loc}}=\Big\{\,\figgg{0.15}{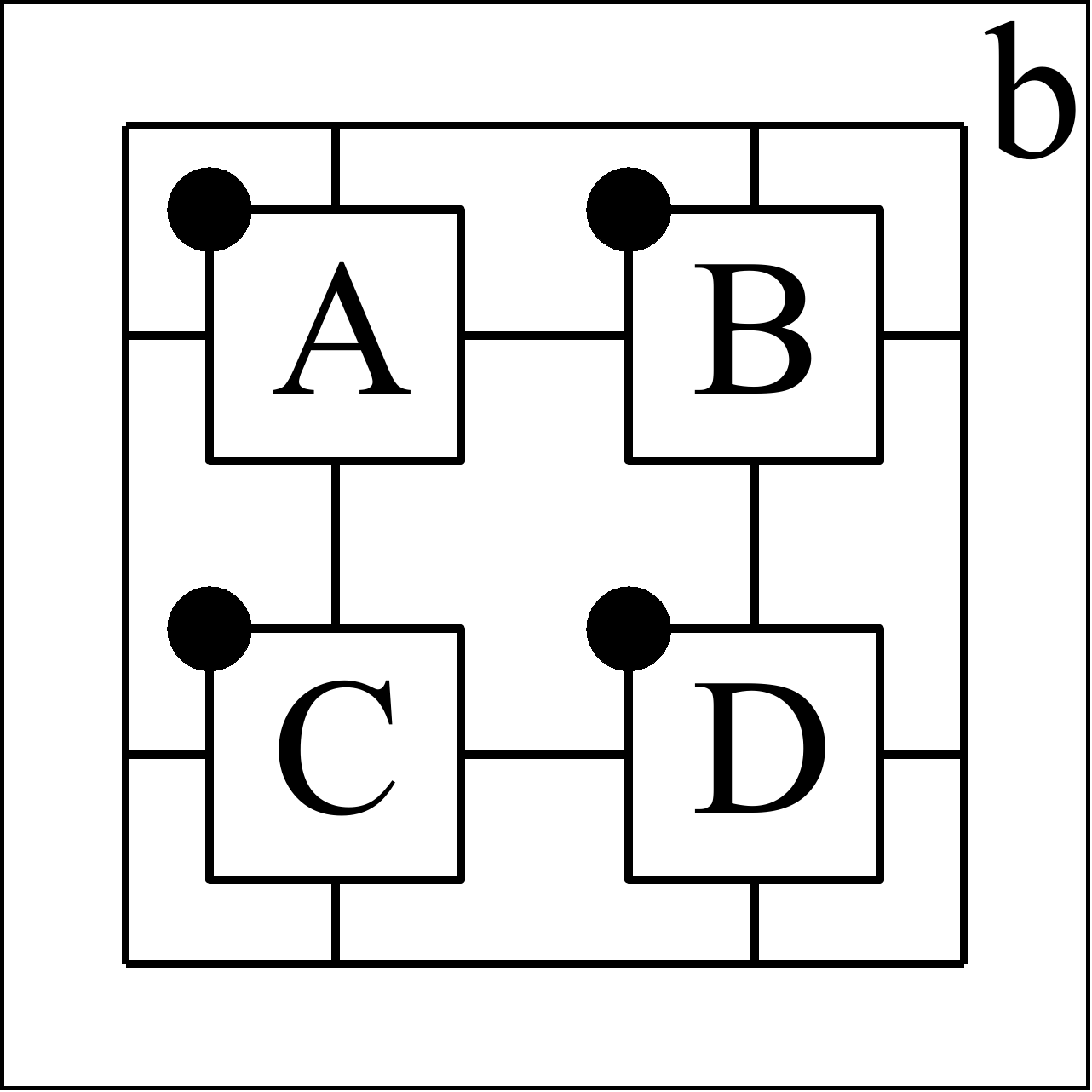}\ ,
    b \hbox{ boundary tensor}\Big\}
\ee
(spanned by all the possible boundary tensors $b$), and summing these
local projectors to construct a global Hamiltonian. The ground space of
this parent Hamiltonian is the intersection of the kernels of the local
projectors.

A PEPS representation of the toric code can be obtained by considering a PEPS
with bond dimension two, and associating the virtual space with the physical
space at every site, $\C^d=(\C^2)^{\otimes 4}$. The tensor $E$ at every
site is then the orthogonal projection onto the space of even spin
configuration in the virtual space,
$E\ket{ijkl}=(1+(-1)^{i+j+k+l})\ket{ijkl}/2$.

{The ground space of the parent Hamiltonian for this PEPS is locally equivalent to the toric code. A detailed treatment of this relationship can be found in \cite{schuch:peps-sym}.}

\emph{Uncle Hamiltonian for the toric code.---}%
Let us now derive the uncle Hamiltonian for the toric code.  This will
be done as for the GHZ state, cf.~Fig.~\ref{fig1}: We perturb the toric code
tensors, derive the corresponding parent Hamiltonian, and take the limit
of vanishing perturbations.  The specific perturbation we consider,
which we denote by $O$, is the projection complementary to $E$, $O=\I
-E$, the projection onto the space of odd spin configurations.

The $2\times 2$-site local Hamiltonian $h_{\mathrm{loc}}^\varepsilon$ is
obtained from Eq.~(\ref{eq:parent2d}) by letting each of the four tensors
be $E+\varepsilon O$. In the limit $\varepsilon\rightarrow 0$, we obtain a 
new projector $h'_{\mathrm{loc}}=\lim h_\mathrm{loc}^\varepsilon$
different from the local projector $h_\mathrm{loc}$ we started with:
The new local Hamiltonian $h'_\mathrm{loc}$ is the projector onto the
orthogonal complement of $E_{22}+O_{22}=\ker h'_{\mathrm{loc}}$, where 
\be
\label{eq:O22-def}
O_{22}=\Big\{\sum_{\mathrm{pos}\,O}\, \figgg{0.15}{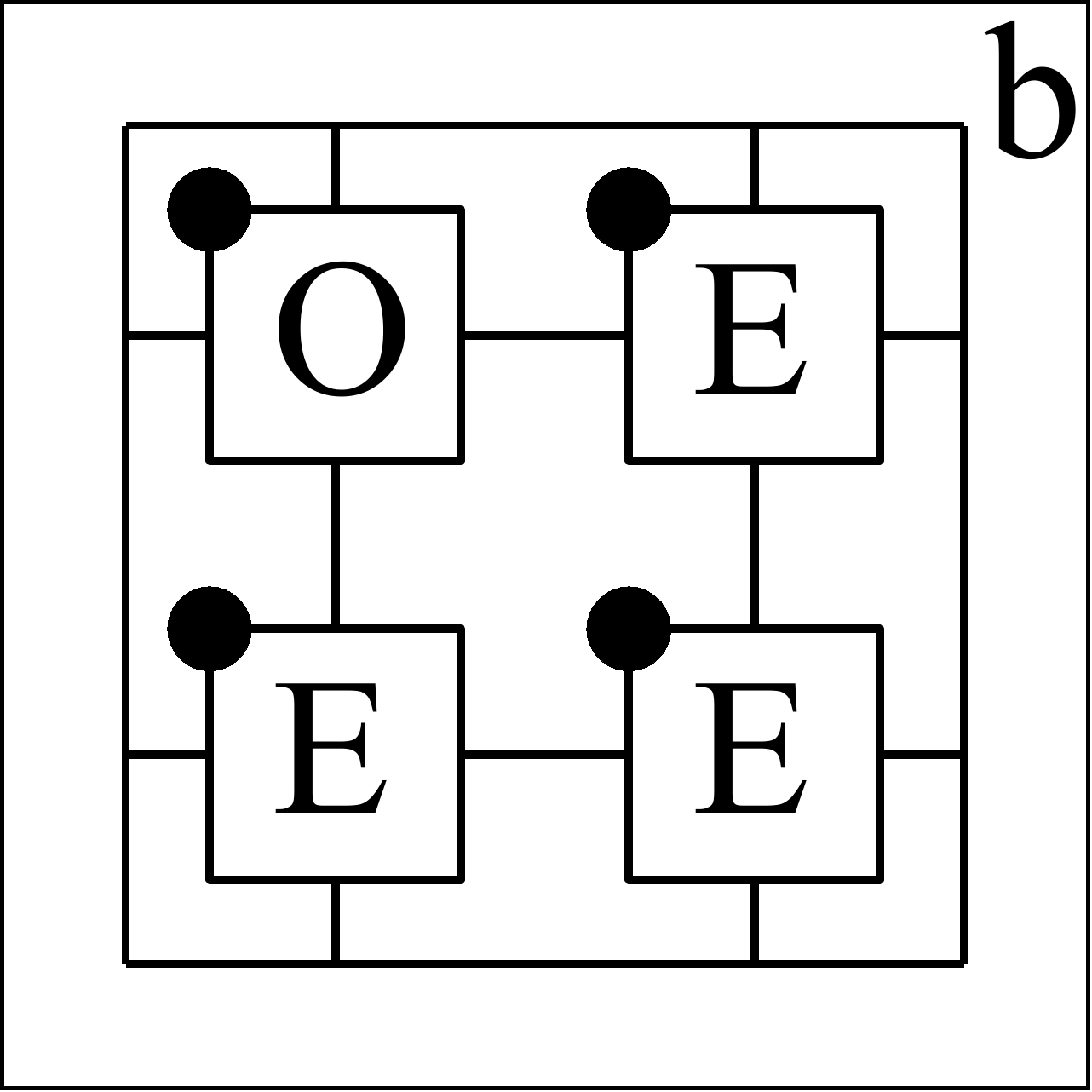}\ , b \hbox{
boundary tensor} \Big\}\ ,
\ee
and the sum runs over the positions which the single $O$ tensor above may
occupy among the four tensors appearing. $E_{22}$ is defined analogouly,
but contains only $E$ tensors.  Note that $E_{22}$ will only be
non-vanishing for even boundary conditions $b$, whereas for $O_{22}$ this
will only be the case for odd boundary conditions.
The space $E_{22}$ plays the role $\spanned\{\ket{000},\ket{111}\}$ did in
the uncle Hamiltonian of the GHZ state, and $O_{22}$ plays the role of
$\spanned\{\ket{0\!+\!1},\ket{1\!+\!0}\}$. Intuitively, while $E_{22}$
only supports states without anyonic excitations, $O_{22}$ allows for
configurations with exactly one anyon which is distributed in a uniform
superposition. As with the domain walls in 1D, the idea is that such
configurations cannot appear in the ground state subspace as anyons come
in pairs, but two excitations are not allowed to meet; however, such
configurations with delocalized anyon pairs will have low energy.

The new uncle Hamiltonian $H'$ is constructed again as the sum over all
$2\times 2$ regions of the local projector $h'_{\mathrm{loc}}$. When
considering an $n\times m$ contractible region $R$, and the sum of the
local projectors acting entirely in this region, one finds that the kernel
of this sum has the same structure as the kernel of a single projector: 
\be
\nonumber
\ker\big(\sum _R h'_{\mathrm{loc}}\big)= \bigcap _R \ker
h'_{\mathrm{loc}}=E_{nm}+O_{nm} \ ,
\ee
with definitions for $E_{nm}$ and $O_{nm}$ similar to
Eq.~(\ref{eq:O22-def}); the detailed proof is given in Appendix A.
However, the $O$ subspace vanishes when considering the whole lattice and
imposing periodic boundary conditions, as those are automatically even
(see Appendix A).
Therefore, the global ground space of the new Hamiltonian is the same as
the ground space of the toric code parent Hamiltonian. 

\emph{Spectrum of the uncle Hamiltonian.---}%
Let us now show that the uncle Hamiltonian for the toric code is gapless
with continuous spectrum in the thermodynamic limit.
As we did with the GHZ uncle Hamiltonian, we will consider a family of low
energy states which are orthogonal to the ground space.  Given any integer
value of $r$, we may take two contractible rectangles $R_1$ and $R_2$ of
size $r\times r$ which are separated by at least two sites.
 We construct a family of unnormalized states $\ket{\phi_r}$ by
placing at these two regions the tensor $O_{rr}$
[cf.~Eq.~(\ref{eq:O22-def})], and setting all remaining tensors to $E$:
\be 
\label{eq:lowenergyvectors}
\ket{\phi_r}=\sum_{ {\mathrm{pos}\ O_1\in R_1}\atop
{\mathrm{pos}\ O_2\in R_2}} \figgg{0.15}{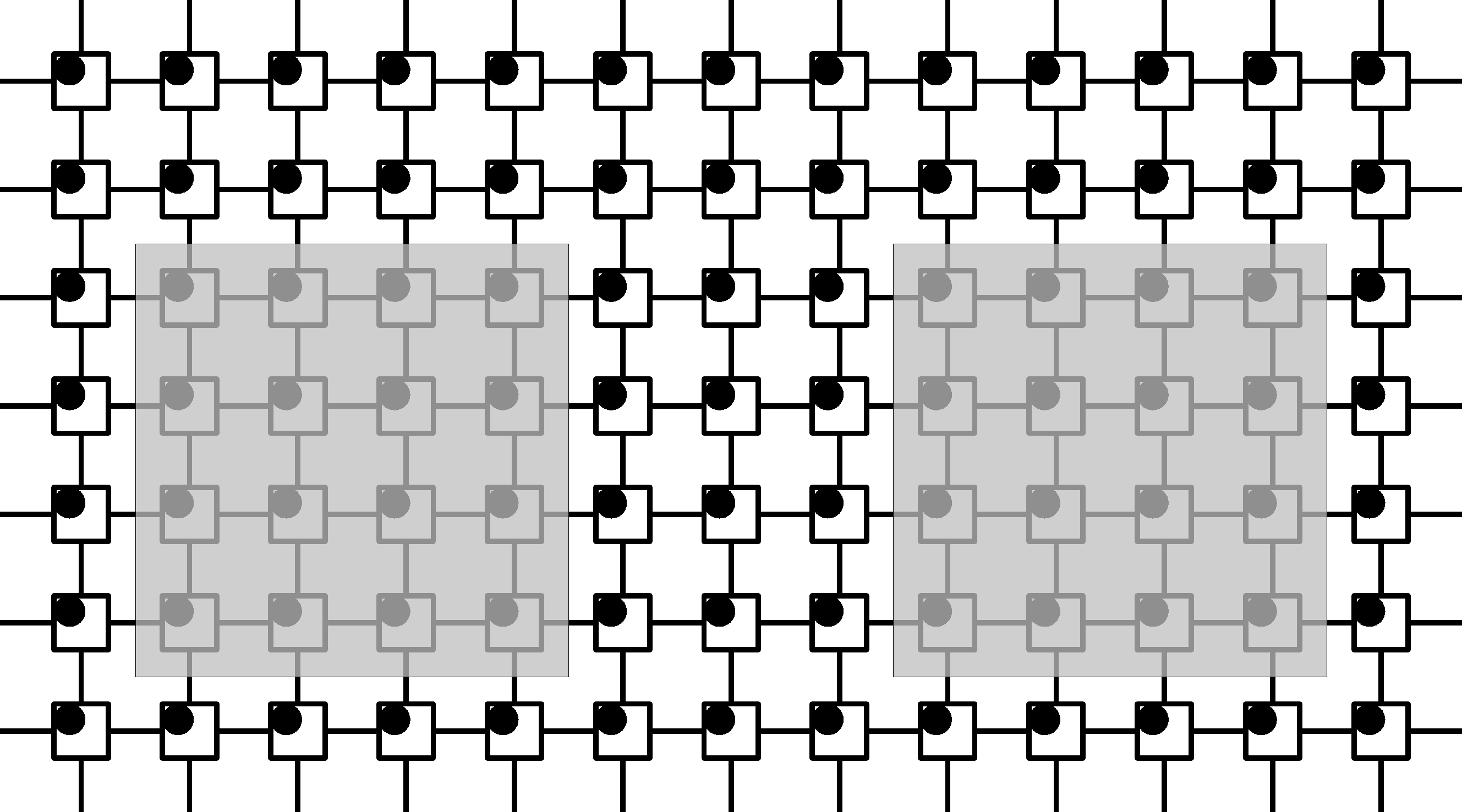} \ .
\ee
This is, each of the gray regions contains exactly one $O$ tensor and
$E$'s otherwise, and the sum runs over the position of the two $O$'s.

The norm of all these summands is the same, say $C$. This value depends
only on the total dimension of the lattice. The norm of any of these
$\ket{\phi_r}$ is $Cr^2$ (since the summands are mutually orthogonal and
there are $r^4$ of them), but only the $h_\mathrm{loc}'$ which overlap
with the boundary of these regions contribute a positive energy. There are only
$8r$ of them, $4r$ acting on the left and $4r$ acting on the 
right region. For each of them at most $2r^2$ summands from (\ref{eq:lowenergyvectors})
add any energy: there are at most two ways $O$ can overlap with the Hamiltonian
term, and the $r^2$ comes from the $O$ in the other region.  Hence
$\bra{\phi_r} H' \ket{\phi_r}\le C^2O(r^3)$, and the
energy $\bra{\phi_r} H' \ket{\phi_r}/ \braket{\phi_r}{\phi_r}$ of these
states decreases as $O(1/r)$.  Altogether, this proves that $H'$ is gapless.

In order to prove that the spectra of these Hamiltonians tend to become dense in the positive real line $\R^+$, we fix one of the dimensions of the
system---let us choose the vertical one---and let the other go to infinity. This results in an MPS-like problem, in which we can take the thermodynamic limit. 

{Since the vertical dimension is fixed -- let us say its value is $N$ -- the regions used to construct the states $\ket{\phi_r}$ from (\ref{eq:lowenergyvectors}) cannot grow indefinitely. We can consider instead similar unnormalized states $\ket{\phi_{r,N}}$, coming from $r\times N$ regions,} to prove the existence of a suitable set of elements in the spectrum $\{\lambda_i\}$ tending to 0, from which it can be shown that any finite sum of these values also lies in the spectrum. These finite sums are dense in $[0,\infty)$, which therefore coincides with the spectrum due to its necessary closedness. 

The same values $\sum \lambda_i$ lie close to eigenvalues of the uncle Hamiltonian for some finite sized --but big enough-- lattices. Hence the spectra of the finite sized uncle Hamiltonians tend be dense in $[0,\infty)$.

The analogue proof for the uncle Hamiltonian of the GHZ is detailed in \cite{fernandez:1d-uncle}, and a sketch of the steps adapted to the toric code can be found in Appendix B.

\emph{Conclusions.---}%
In this paper, we have used the framework of PEPS to study different ways
in which strongly correlated quantum systems can appear as ground states
of local Hamiltonians.  In particular, we have introduced the 
\emph{uncle Hamiltonian} of a PEPS, which contrasts with the usually
considered parent Hamiltonian.
 The uncle Hamiltonian is
obtained by perturbing the PEPS tensors, computing the corresponding
parent Hamiltonian, and then taking the perturbation to zero.  As parent
Hamiltonians of systems with degenerate ground states are not robust under
perturbations of the tensors, the resulting uncle Hamiltonian behaves very
different from the parent Hamiltonian:
While the
parent and the uncle Hamiltonian share the same ground state space, the
uncle is gapless with a continuous spectrum in the thermodynamic
limit, thus behaving very differently.  We have demonstrated our approach
with Kitaev's toric code: The resulting uncle Hamiltonian has the toric
code state as its ground state, however, it is gapless and thus does not
yield a topologically protected system.  This both demonstrates the
ambiguity in the association of PEPS with local Hamiltonians, and the
subtleties one has to take care of when identifying topological order from
the ground state rather than the properties of the interaction.

\emph{Acknowledgments.---}%
We specially thank Bruno Nachtergaele for useful comments and discussions.
This work has been partially funded by the Spanish grants I-MATH,
MTM2008-01366, S2009/ESP-1594 and QUITEMAD, the European project QUEVADIS, the Gordon and Betty Moore Foundation through Caltech's
Center for the Physics of Information, the NSF Grant No.\ PHY-0803371, and
the ARO Grant No.\ W911NF-09-1-0442.  We also acknowledge the hospitality
of the Centro de Ciencias de Benasque Pedro Pascual and of the Perimeter Institute, where part of this
work was carried out.


\onecolumngrid

\appendix

\section*{Appendix A: Ground space of the uncle Hamiltonian for the toric
code}

In this appendix, we derive the structure of the ground space of the toric
code uncle Hamiltonian.  Recall first how parent and uncle Hamiltonians
are constructed. Being $E$ ($O$) the orthogonal projection on 
$(\C^2)^{\otimes 4}$ onto the subspace of even (odd) spin
configurations, we can consider the spaces 
\[
\nonumber E_{22}=\Big\{\, \figgs{0.15}{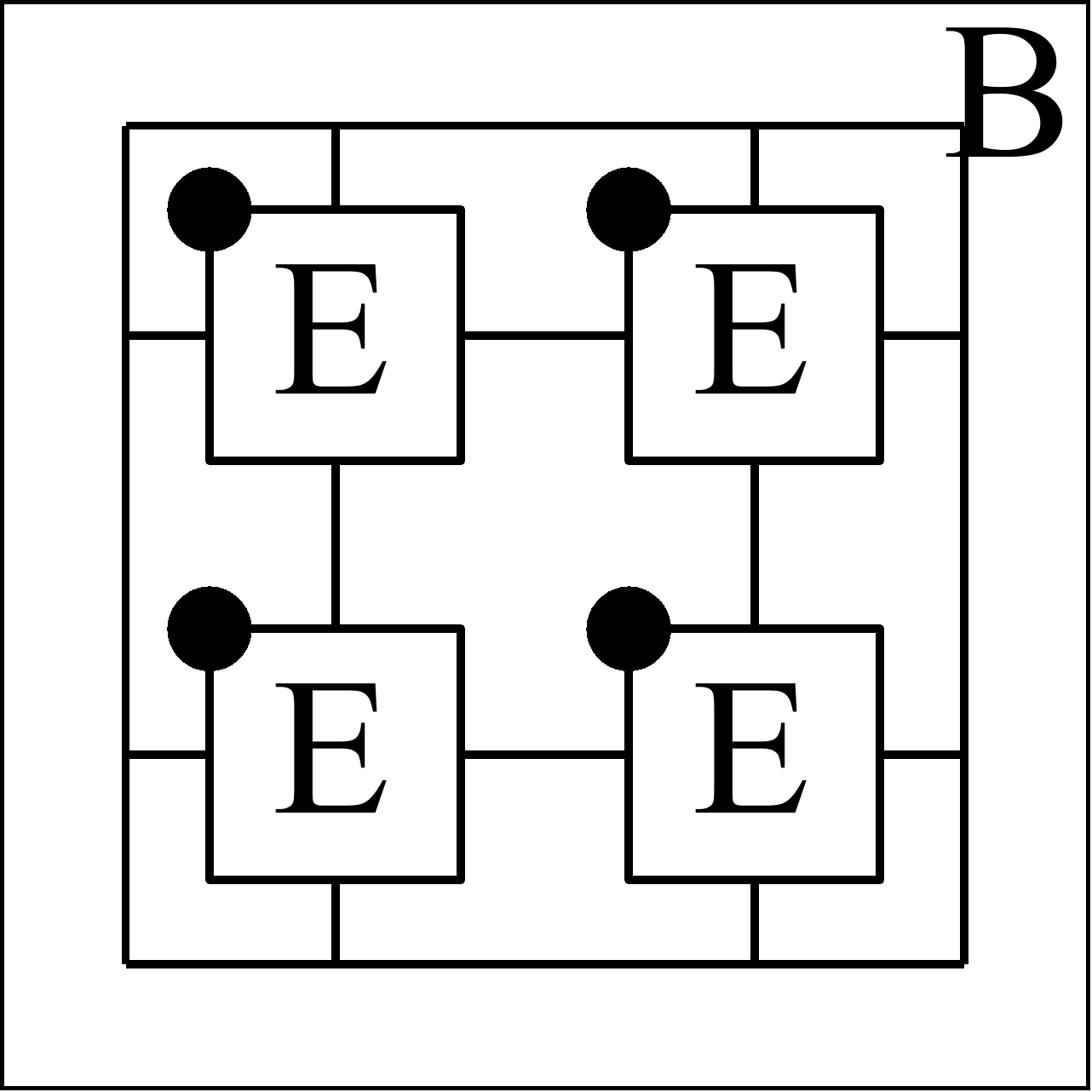}\ ,\, B \hbox{ boundary condition}
\Big\}
\]
and
\[
\ O_{22}=\Big\{\sum_{\mathrm{pos}\ O}\, \figgs{0.15}{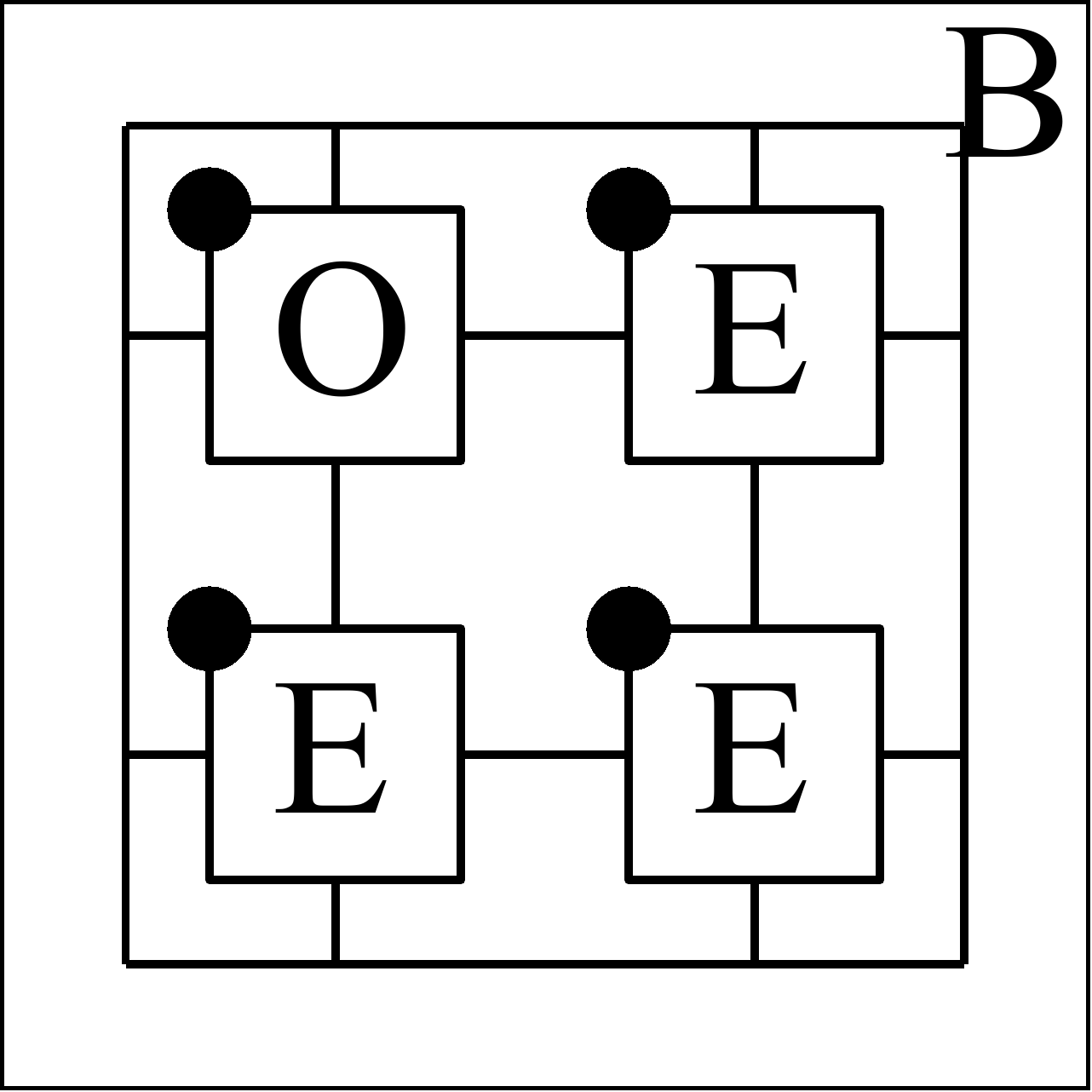}\ ,\,B \hbox{ boundary
condition} \Big\}\ .
\]
The parent Hamiltonian is then constructed as the sum over every $2\times2$ square sublattice of the projection $h_{\mathrm{loc}}$ onto the orthogonal complement of $E_{22}$ (that is, $\ker h_{\mathrm{loc}} =E_{22}$). The uncle Hamiltonian is constructed in the same way, but its local Hamiltonian $h'_{\mathrm{loc}}$ has as kernel the space $E_{22}+O_{22}$.

The structure of the ground space of the parent Hamiltonian can be found
in \cite{schuch:peps-sym}. We will follow here the same steps in deriving
the ground state subspace of the uncle Hamiltonian, allowing the reader
interested in further details to find them in~\cite{schuch:peps-sym}.

We will prove in Proposition 1 that the intersection of the kernels of a
family of local Hamiltonians $h'_{\mathrm{loc}}$ effectively acting on a
given sublattice with dimension $n\times m$, which we will call $S_{nm}$,
keeps having the same structure. It is the vector space:

$$S_{nm}=E_{nm}+ O_{nm}$$
where
\begin{align*}
E_{nm}&=\spanned\Big\{\,
    \figgs{0.15}{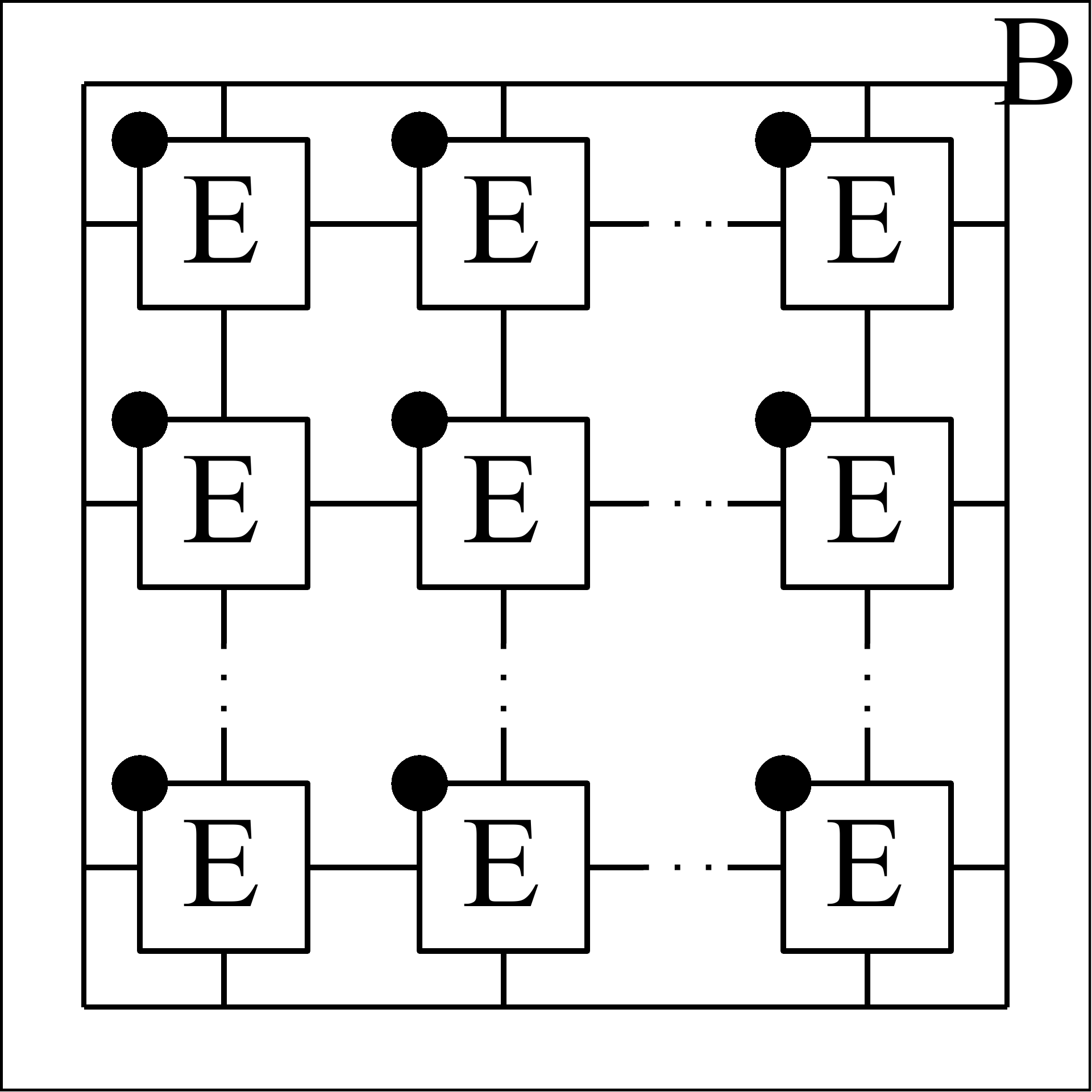},\  B \hbox{ boundary condition}\Big\}\ ,
\\[2ex]
O_{nm}&=\spanned\Big\{
    \sum_{\mathrm{pos}\,O}\,\figgs{0.15}{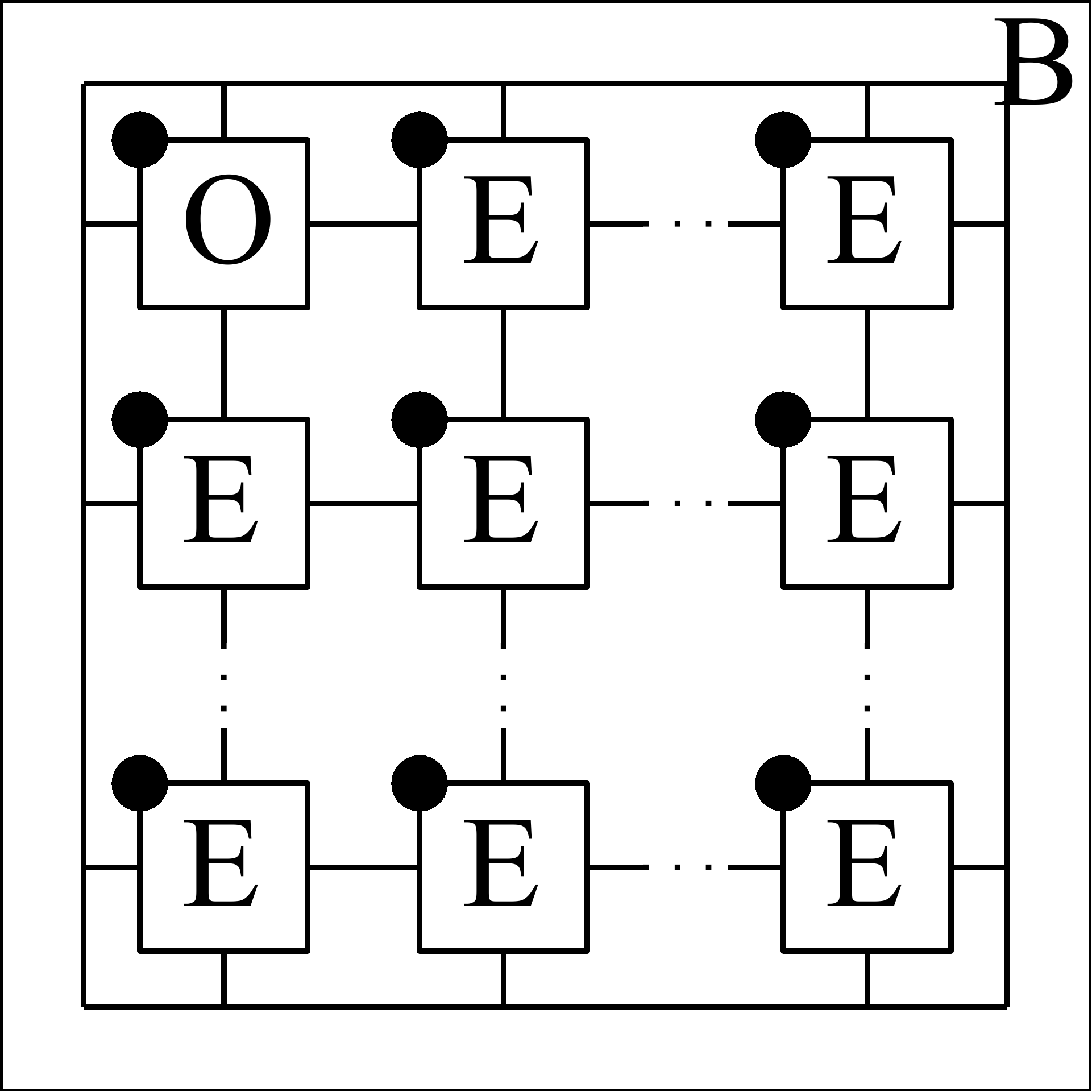}\,,\ B \hbox{ boundary
    condition }\Big\}\ .
\end{align*}
Let us note that in $E_{nm}$ only even boundary conditions give rise to
non-zero vectors, and in $O_{nm}$ only odd boundary conditions do so.
However, as we show in Proposition 2, the $O$ summand disappears when
imposing periodic boundary conditions to the full $N\times M$ lattice,
and the ground space of the uncle Hamiltonian is exactly the same as the
ground space of the parent Hamiltonian.

Let us first prove that the intersection of the kernels of the
$h_\mathrm{loc}'$ is indeed described by $S_{nm}$. The following
proposition serves as the first step in an induction over $n$ and $m$.

\begin{proposition}[Intersection property] Given a $2\times3$ lattice,
        $S_{22}\otimes \C^{2^8} \cap  \C^{2^8}\otimes S_{22}=S_{23}$.
\end{proposition}

\noindent{\bf Proof.} Let $\ket{\phi}$ be an unnormalized vector in
$S_{22}\otimes \C^{2^8} \cap  \C^{2^8}\otimes S_{22}$. This vector can be
written in two different ways:

\begin{equation}
\label{intersection1}
\ket{\phi}
= \ \figgs{0.15}{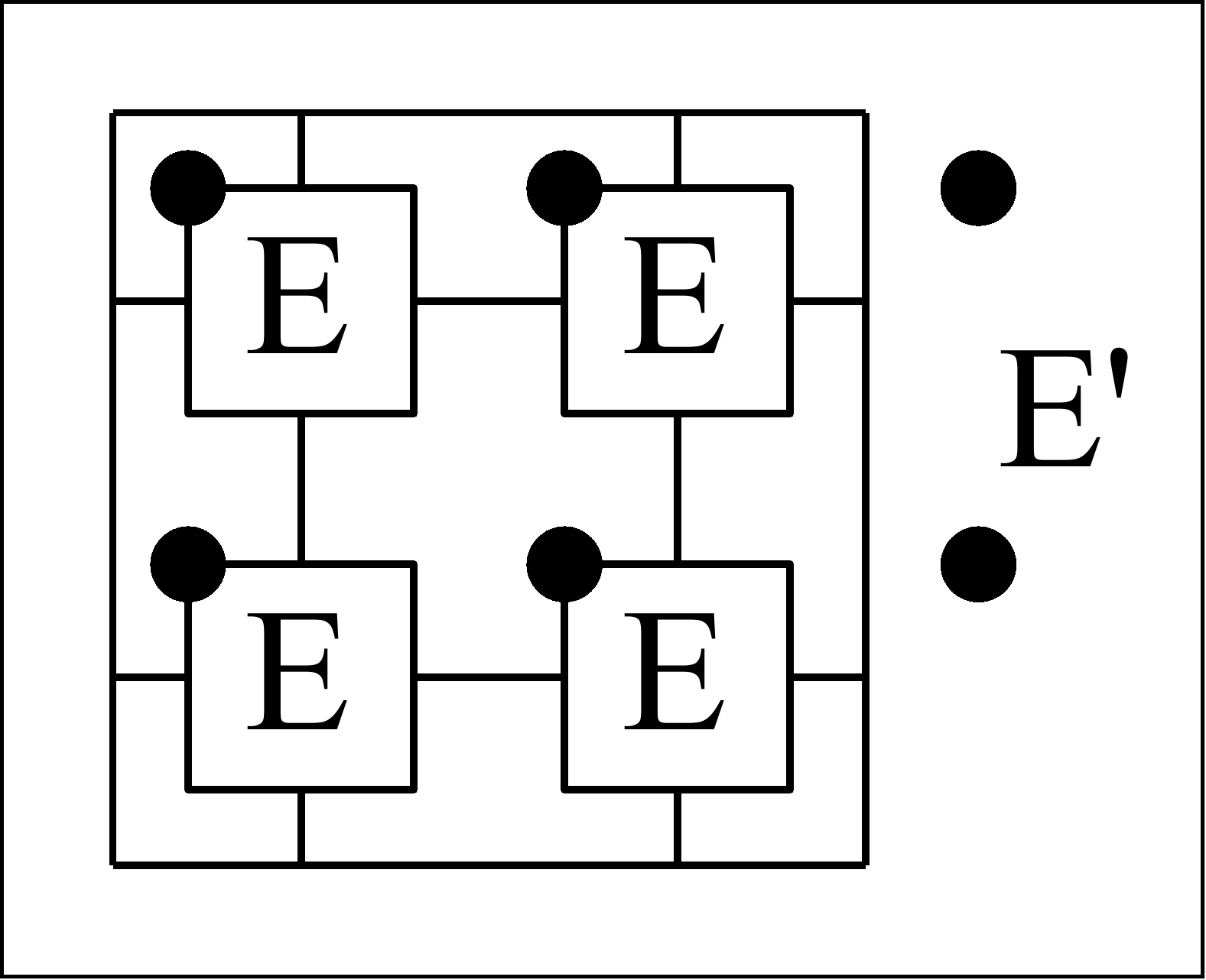}+\sum_{\mathrm{pos}\,O}\,\figgs{0.15}{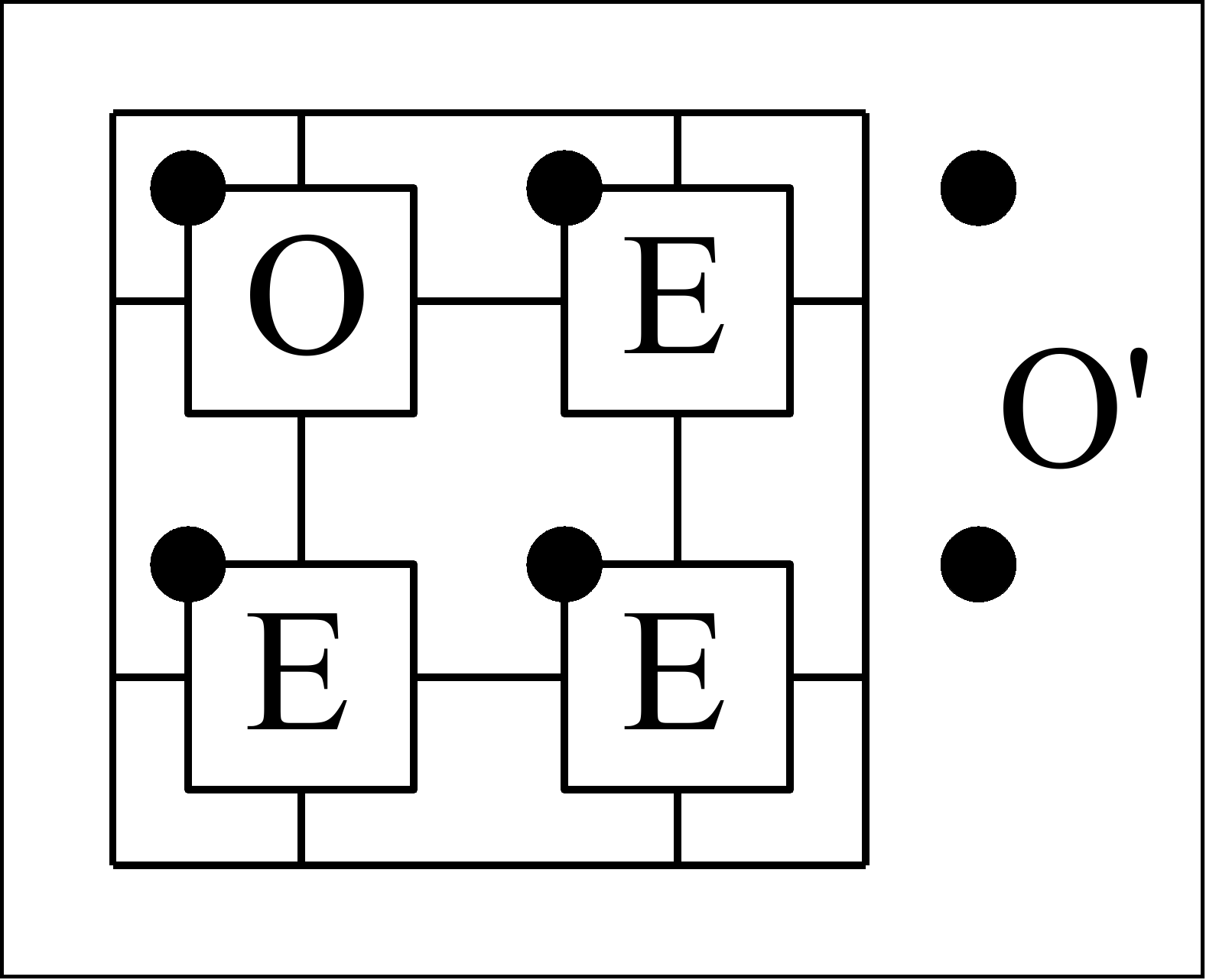}
\quad = \quad \figgs{0.15}{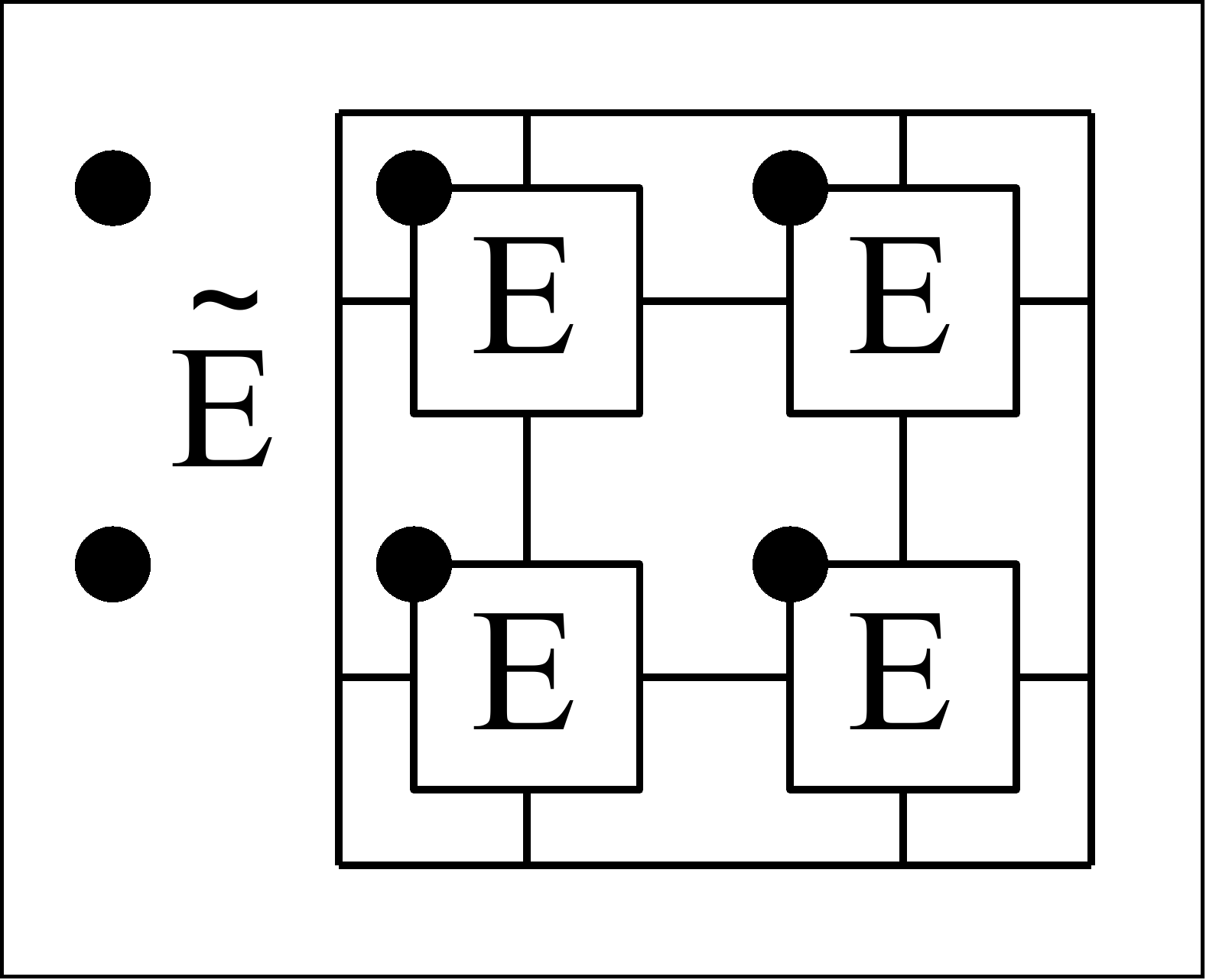}+\sum_{\mathrm{pos}\,O}\,\figgs{0.15}{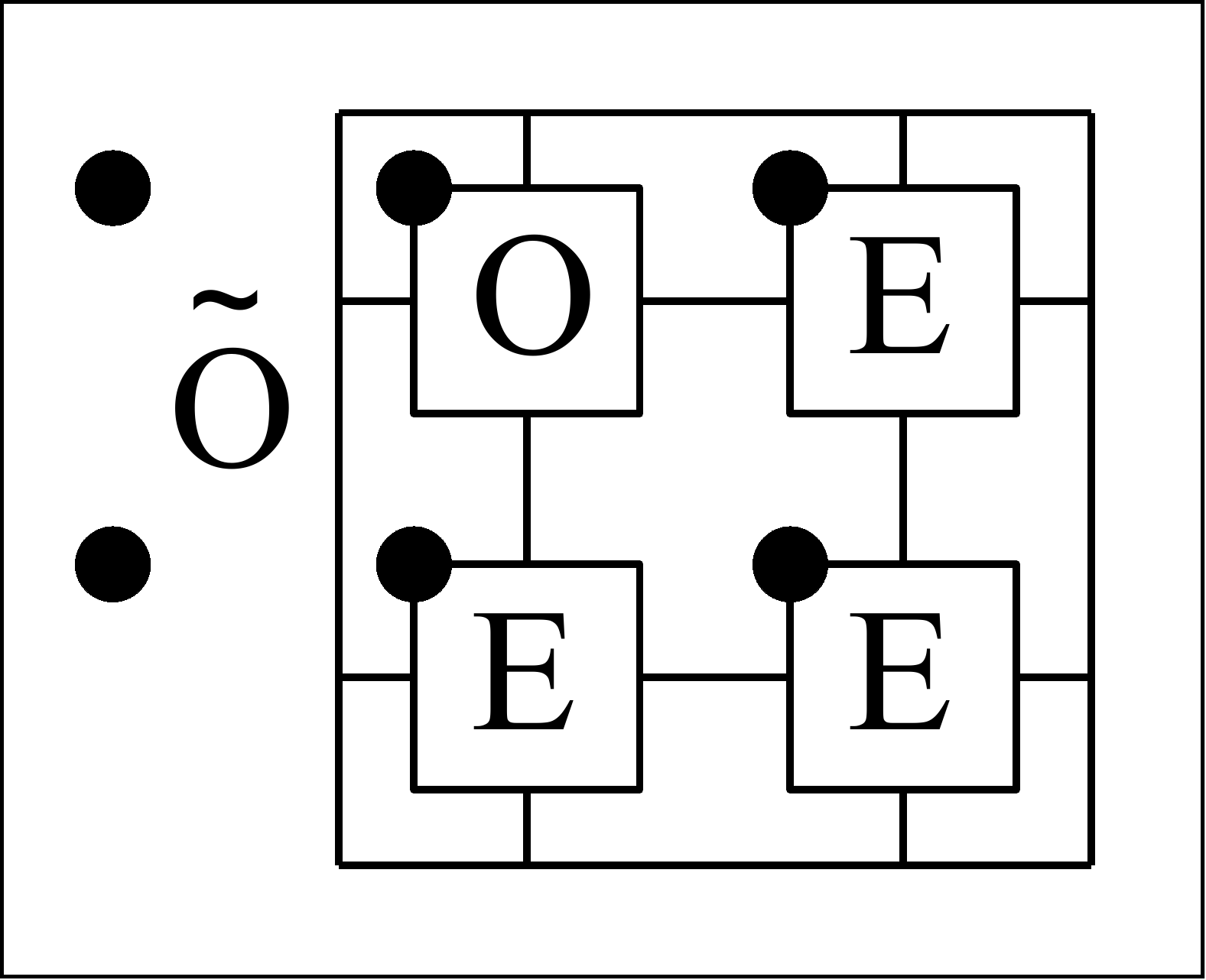}
\end{equation}
W.l.o.g.\ we can assume that the boundary conditions given by $E'$ and
$\tilde{E}$ are always even, and those given by $O'$ and $\tilde{O}$ are
always odd.

We will now perform the projection 
\[
\figgs{0.15}{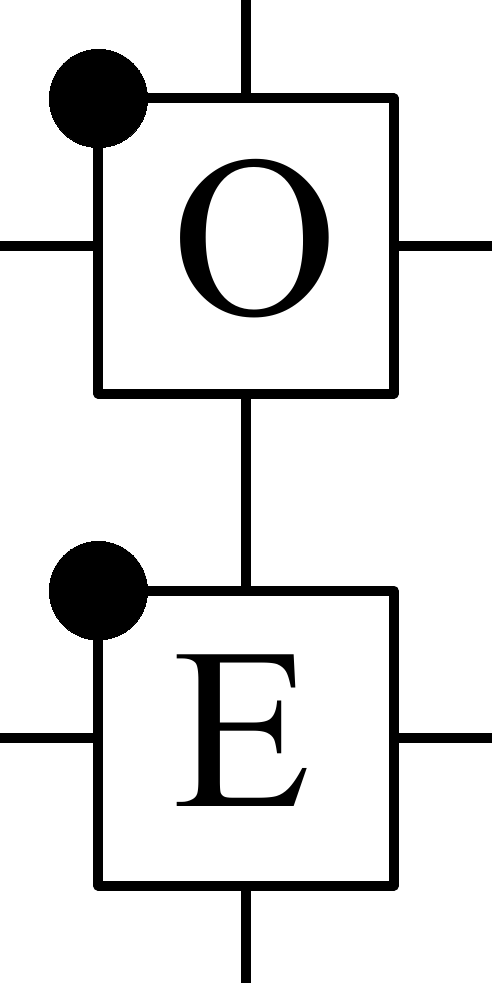} + \figgs{0.15}{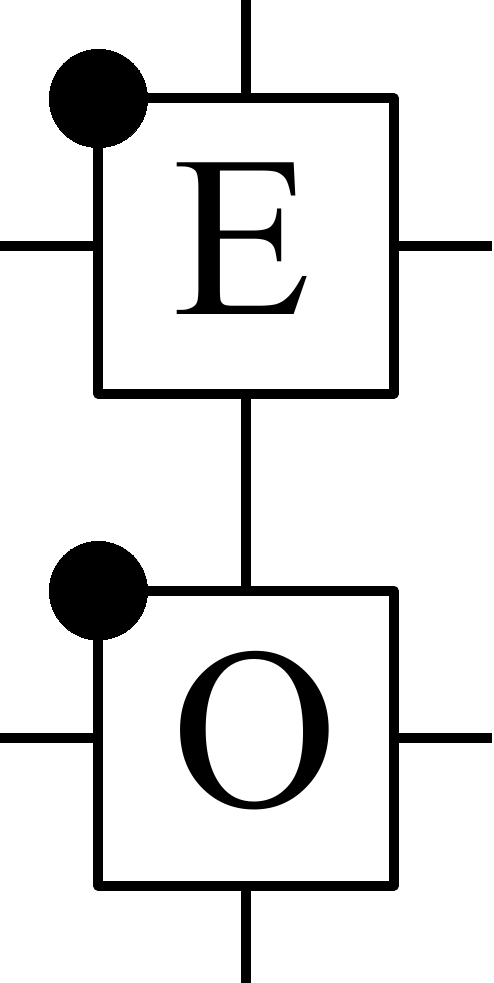}
\]
on the physical levels in the second column.  As different configurations
of $E$'s and $O$'s are orthogonal, this exactly selects this pattern
in the second column, and we obtain the equality
\begin{equation}
\label{intersection2}
\figgs{0.15}{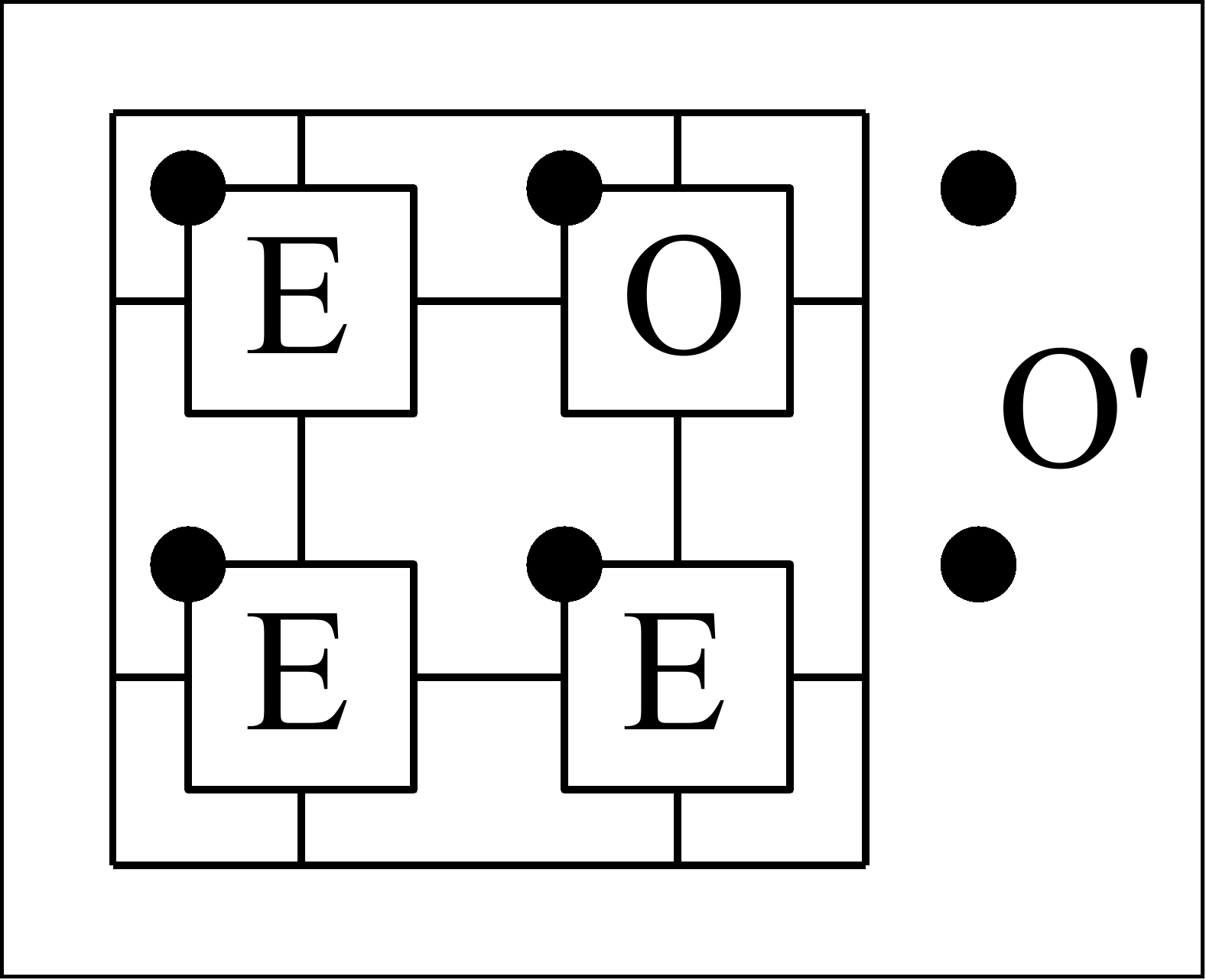}\ +\ \figgs{0.15}{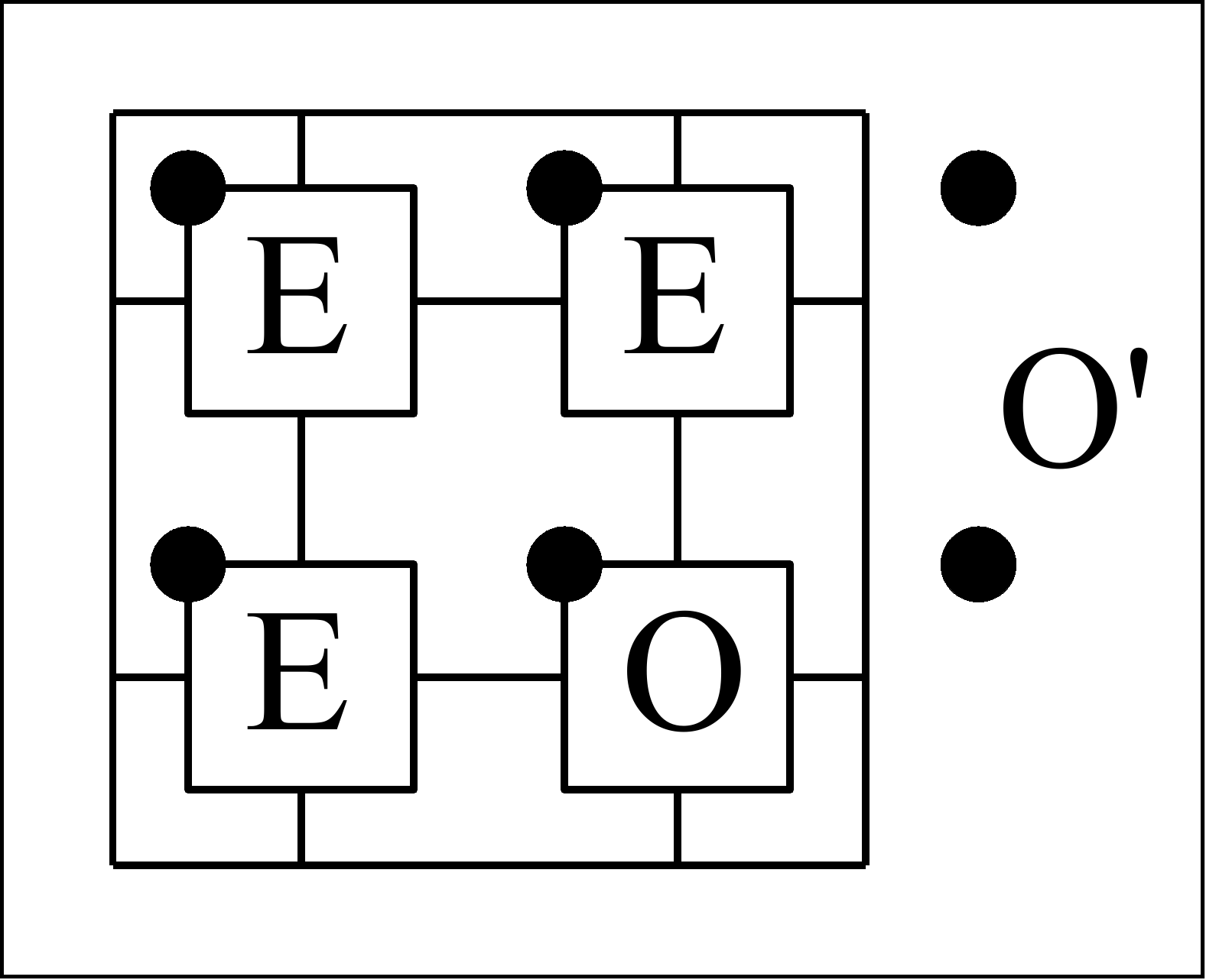}
\quad=\quad \figgs{0.15}{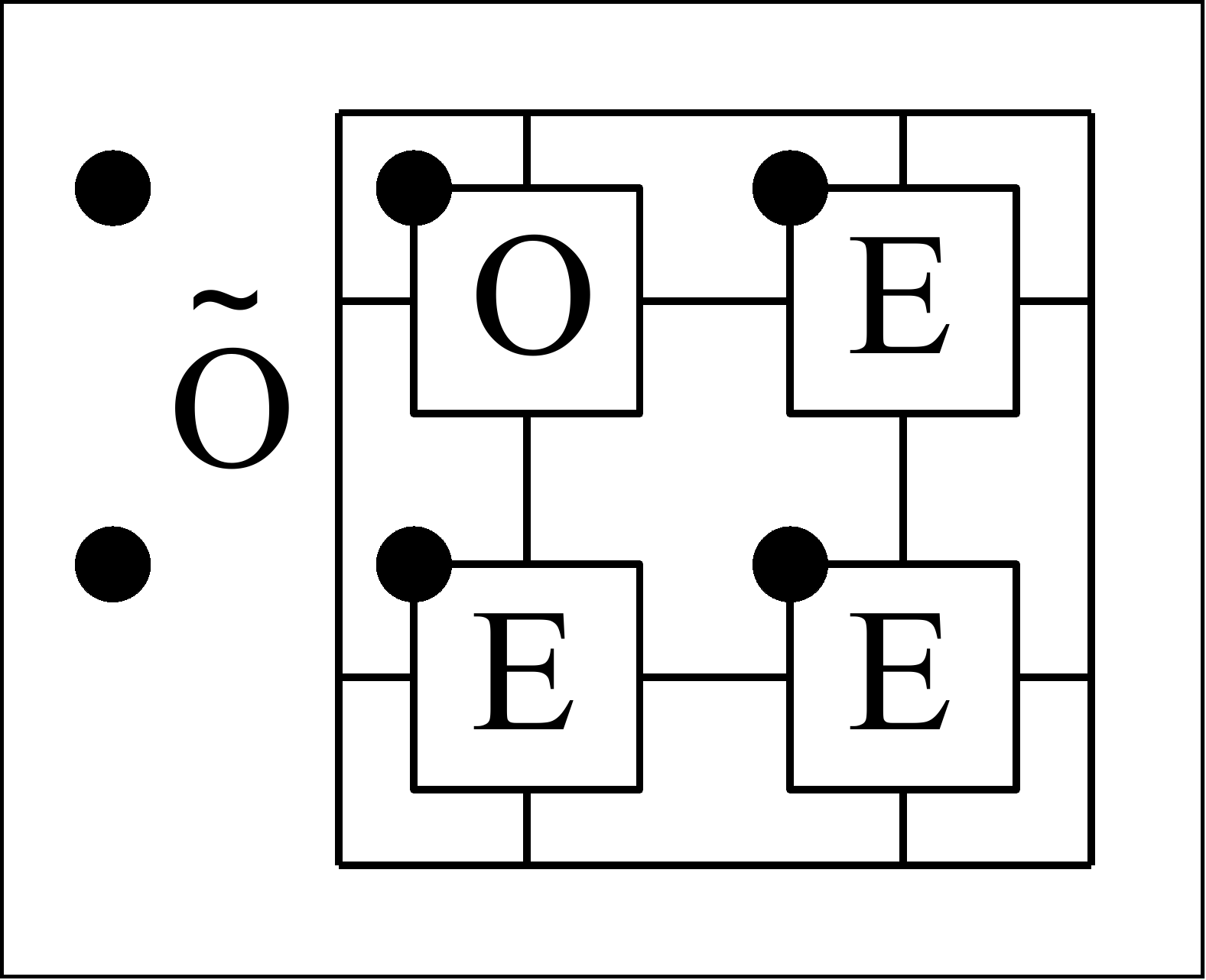}\ + \figgs{0.15}{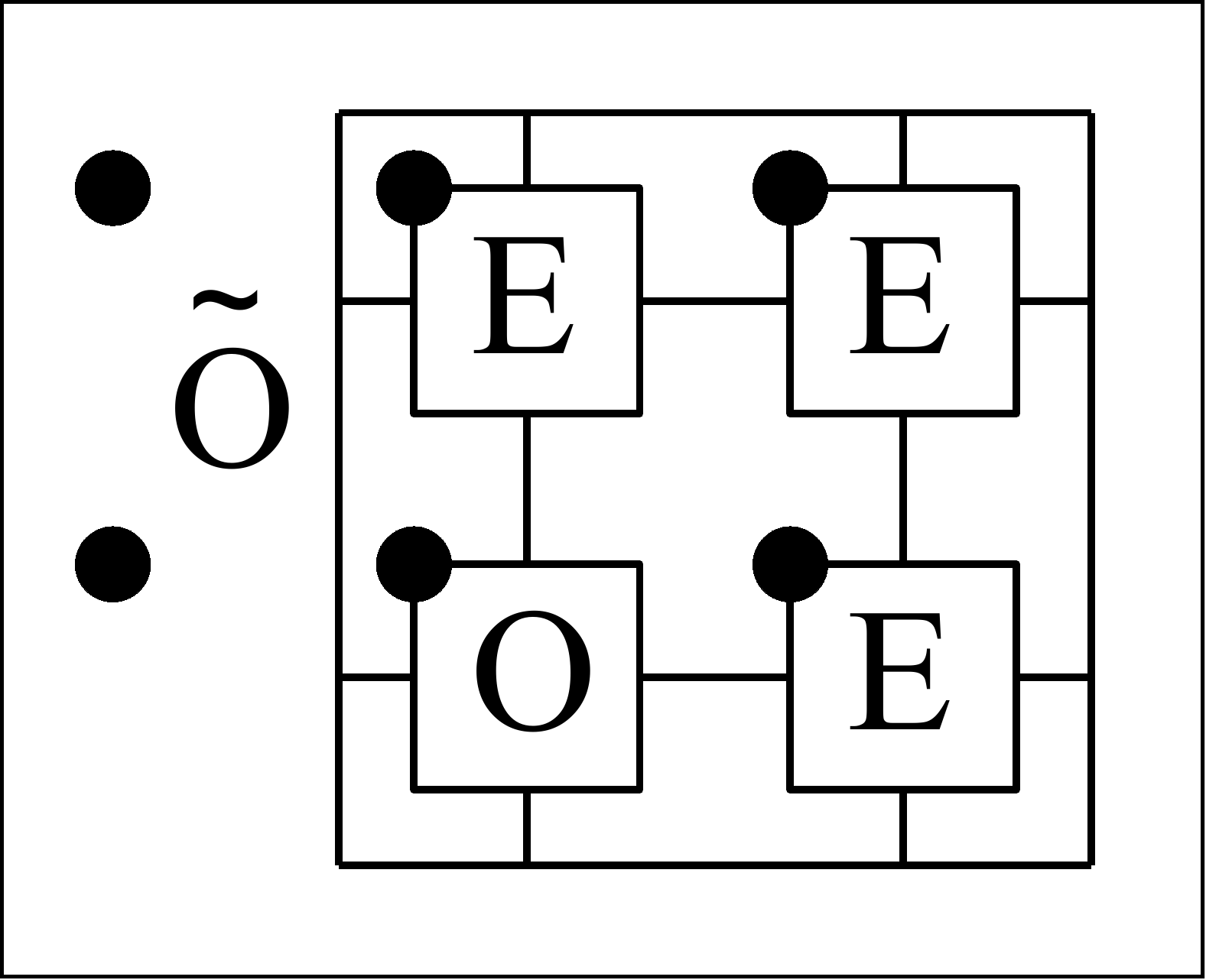}\ .
\end{equation}
In order to infer the structure of $O'$ and $\tilde O$, we will now
project either the first or the third column onto
\[
\figgs{0.15}{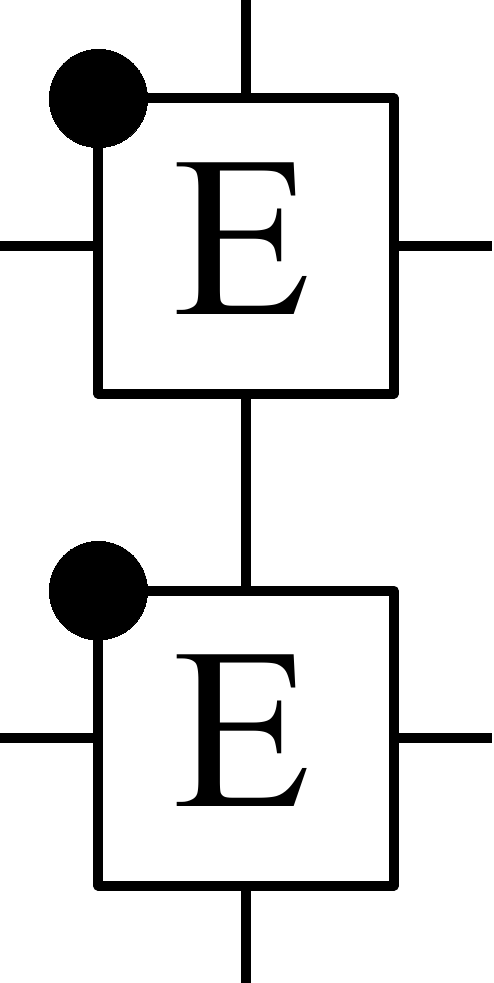}\quad ,
\]
and use the fact that \emph{i)} $O'$ and $\tilde O$ have
odd parity and $\emph{ii)}$ the resulting tensor network of $E$'s and
$O$'s is equivalent to a projection onto the odd parity subspace.
By projecting the first row, we find that
\begin{equation}
\label{eq:intersection2p5}
\figgs{0.15}{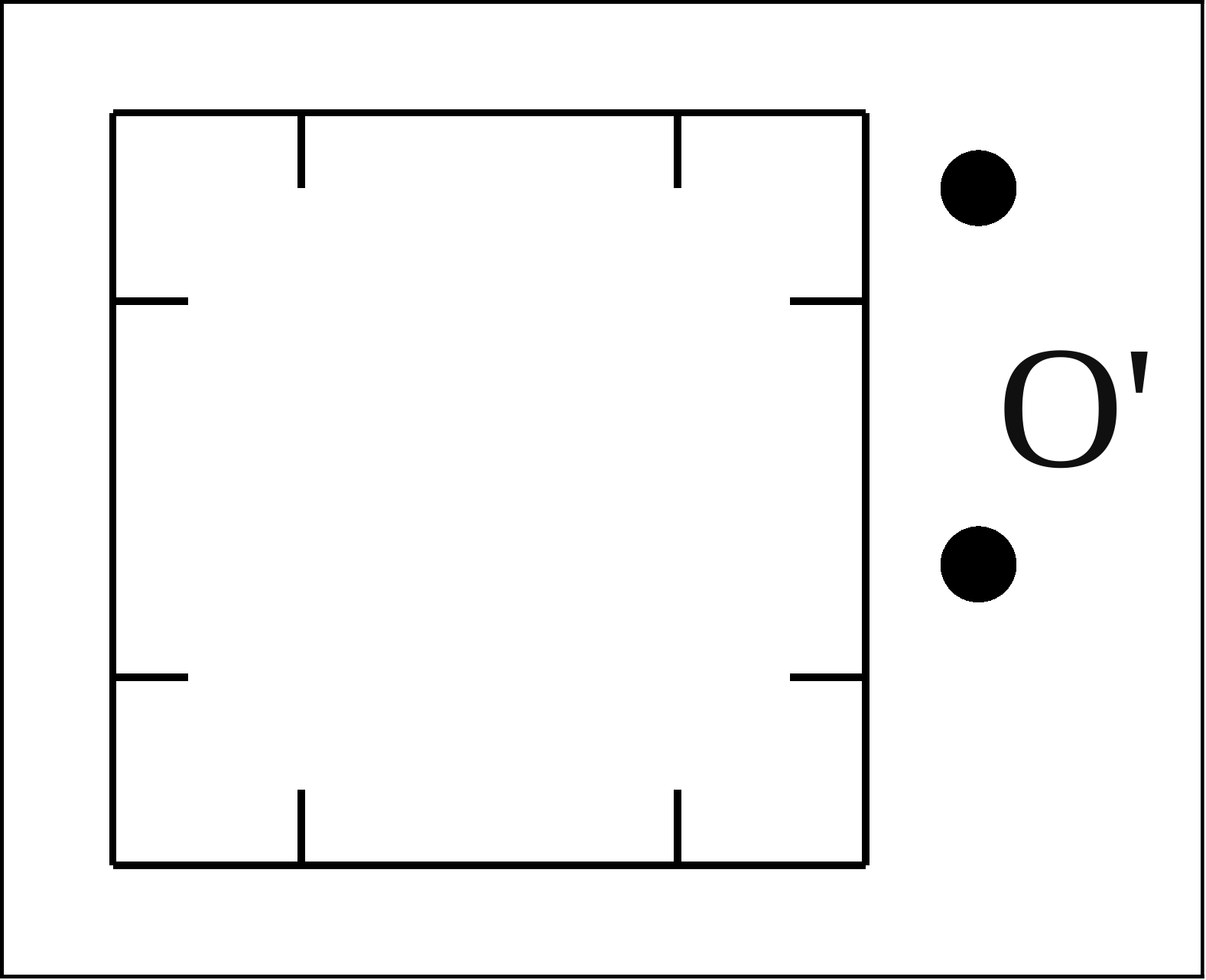}\quad = 
\figgs{0.15}{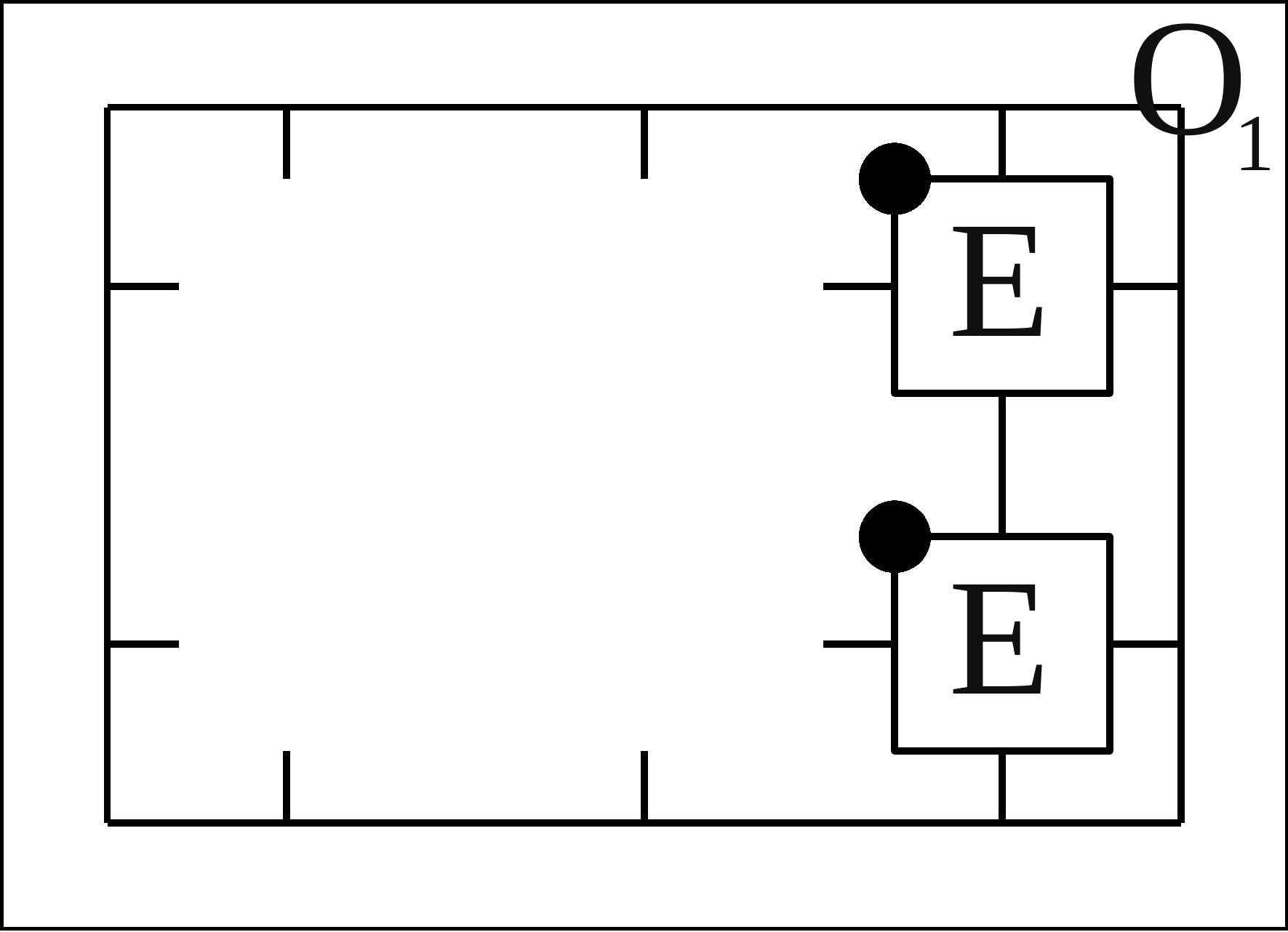}\ ,
\end{equation}
and by projecting the third row, we obtain a corresponding equation for
$\tilde O$ with a boundary $O_2$. Re-substituting in
(\ref{intersection2}), we find that $O_1=O_2$; moreover, the new boundary
condition has odd parity.
Substituting Eq.~(\ref{eq:intersection2p5}) and its analog for $\tilde O$
back into Eq.~(\ref{intersection1}), we obtain 
\begin{equation}
\label{intersection3}
\ket{\phi}
=\ \figgs{0.15}{Eprima}+\!\!\!\sum_{\mathrm{pos}\,O\in\figgs{0.15}{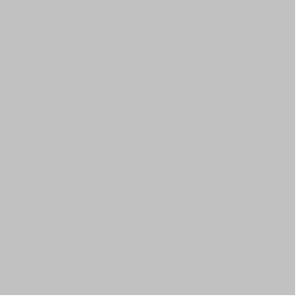}}\,
    \figgs{0.15}{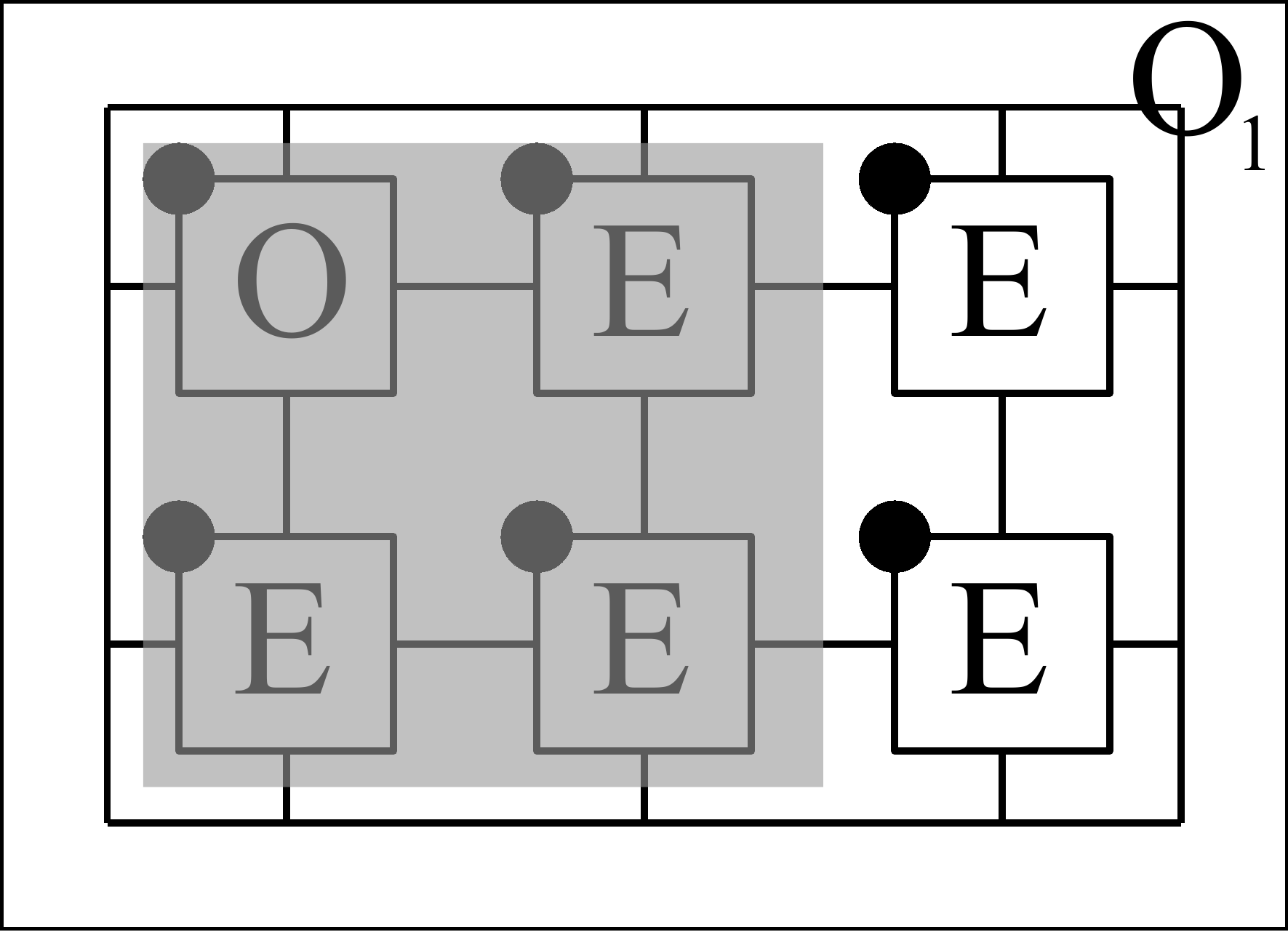}
\quad =\quad
\figgs{0.15}{Etilde}+\!\!\!\sum_{\mathrm{pos}\,O\in\figgs{0.15}{gris}}\,
    \figgs{0.15}{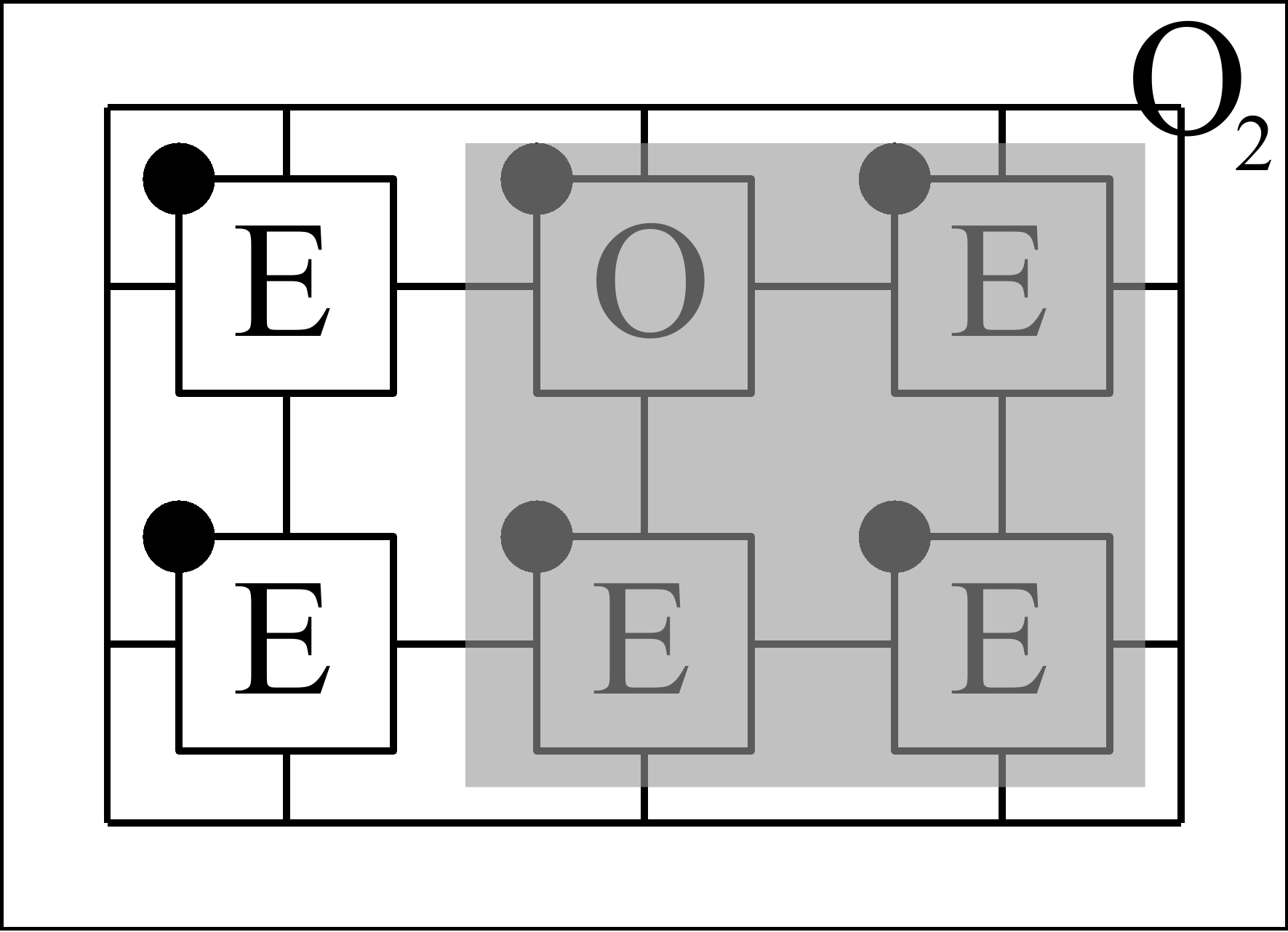}\ ,
\end{equation}
where the sums run over all positions of the $O$ tensor inside the gray
regions.  

We now use the same trick to also infer the structure of $E'$ and $\tilde
E$: We apply the projection 
\[
\figgs{0.15}{EE} + \figgs{0.15}{OE} + \figgs{0.15}{EO}
\]
in either the first or the third column of Eq.~(\ref{intersection3});
after re-substituting the resulting conditions,
 we find that
\begin{align*}
\ket{\phi}&=\ 
    \figgs{0.15}{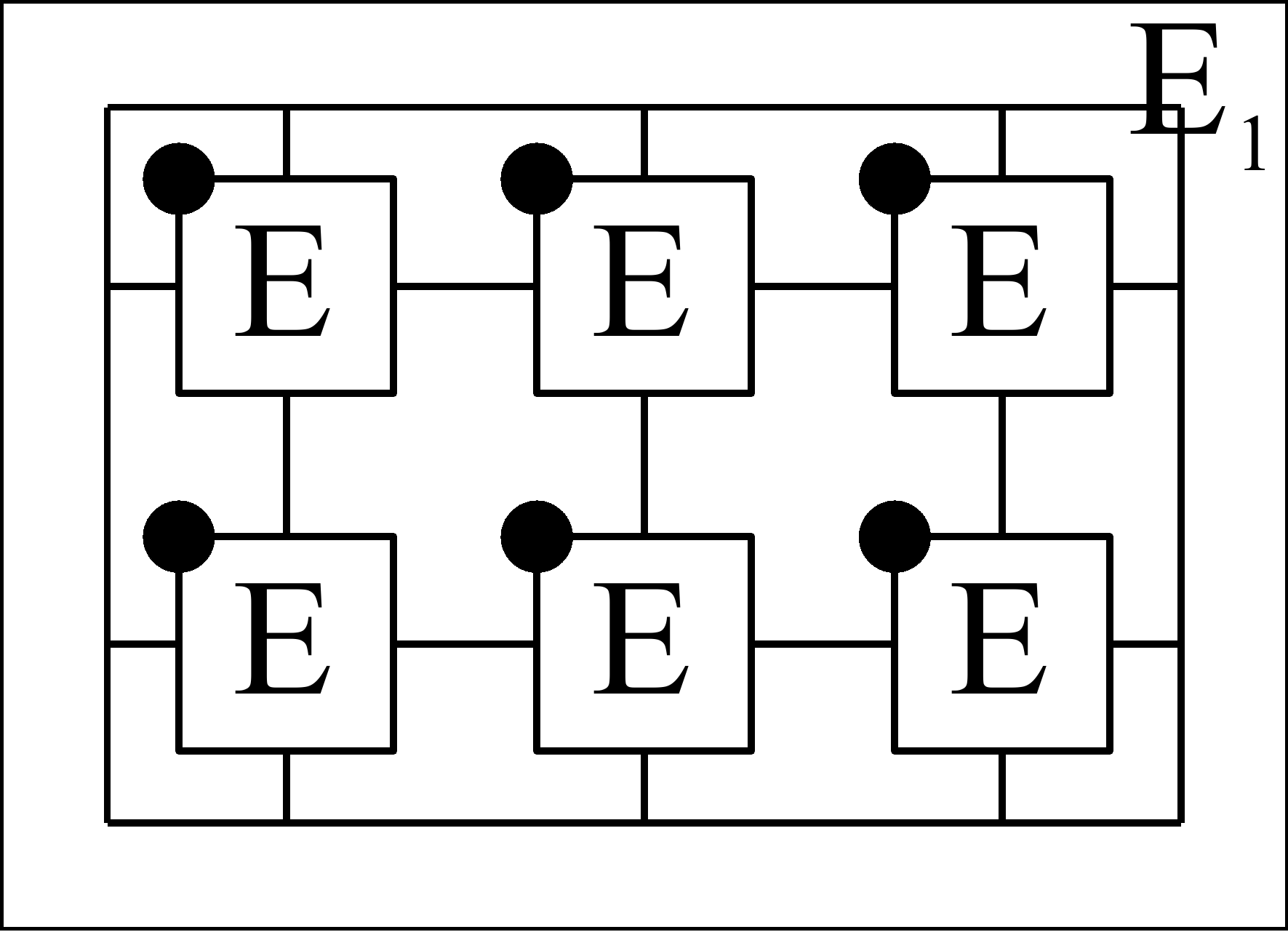}\
    +\sum_{\mathrm{pos}\,O\in\figgs{0.15}{gris}}\,\figgs{0.15}{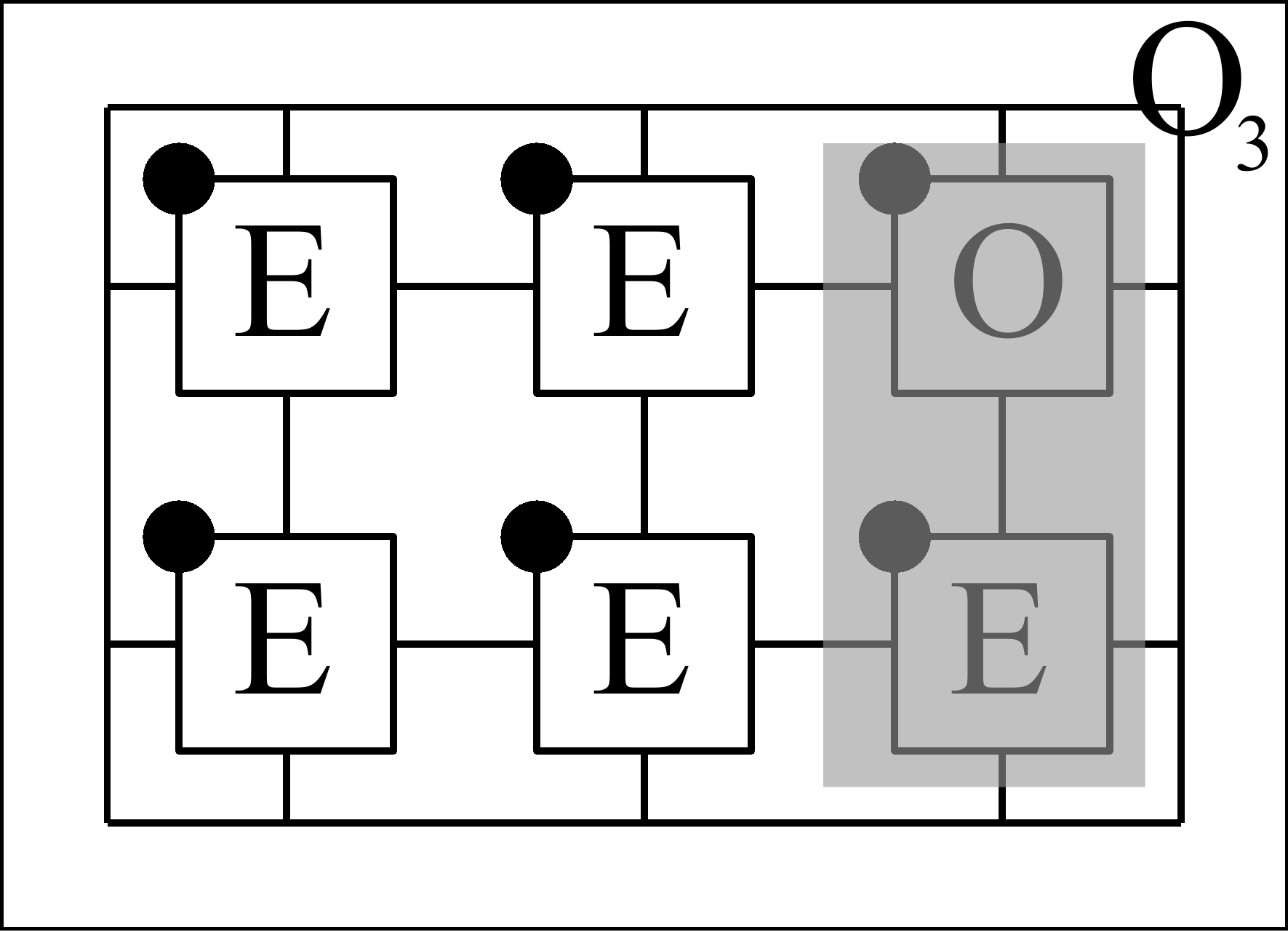}\ 
    + \sum_{\mathrm{pos}\,O\in\figgs{0.15}{gris}}\,\figgs{0.15}{Int3-2} 
\\[2ex]
&=\ \figgs{0.15}{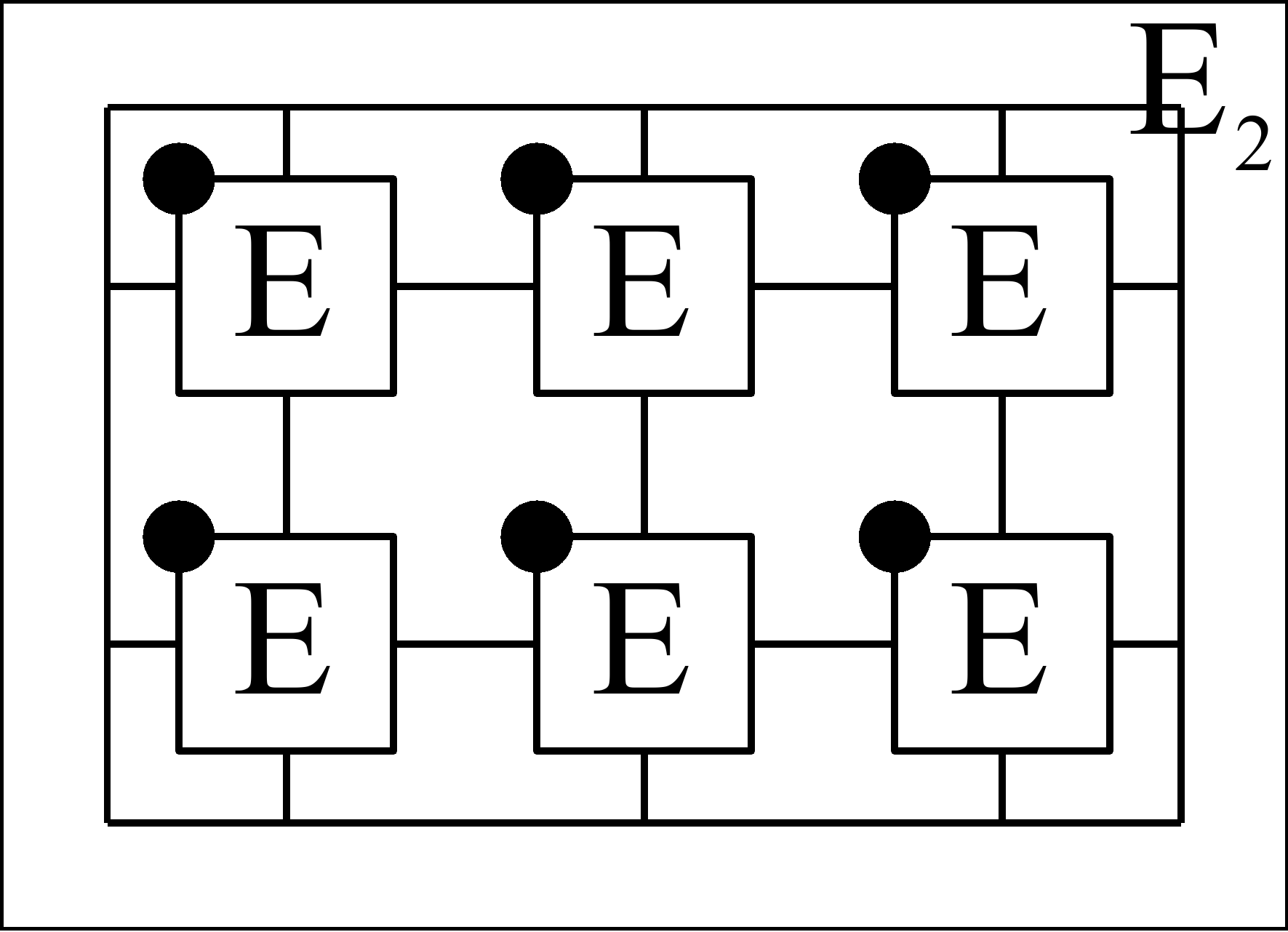}\ + 
\sum_{\mathrm{pos}\,O\in\figgs{0.15}{gris}}\,\figgs{0.15}{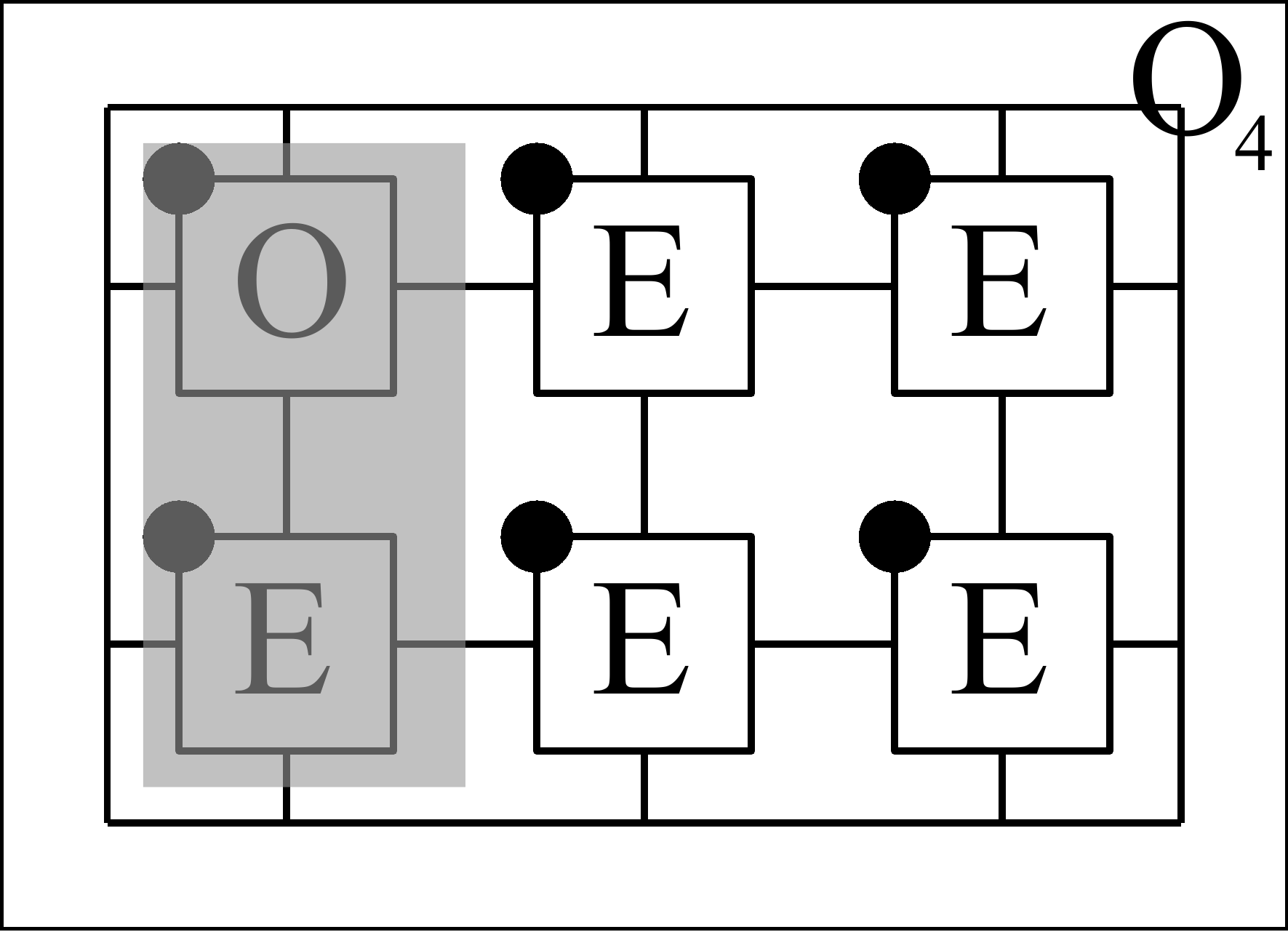}
\ + \sum_{\mathrm{pos}\,O\in\figgs{0.15}{gris}}\,\figgs{0.15}{Int3-4}\ .
\end{align*}
By matching equal patterns of $E$'s and $O$'s, we can easily check that
$E_1=E_2$, $O_1=O_4$, and $O_2=O_3$.  Thus, there exist unique even and
odd boundary conditions $B_E=E_1=E_2$ and $B_O=O_1=O_2=O_3=O_4$ which
describe the state $\ket{\phi}$ as an element from $S_{23}$.
\hspace*{\fill}$\square$

Using this argument inductively, we can indeed prove for any contractible
rectangle of size $n\times m$ (or in fact any contractible region) that
$S_{nm}$ is equal to the intersection of the kernels of the local
Hamiltonians $h'_\mathrm{loc}$ which act inside the region.

\begin{proposition}[Closure property] The ground space of the uncle
        Hamiltonian coincides with the ground space of the parent
        Hamiltonian.
\end{proposition}

\noindent{\bf Proof.} 
Exploiting the $\sigma_z$ symmetry of $E$ and $O$ tensors, we can prove that for a state to lie in the kernel of every $h'_{\mathrm{loc}}$, and therefore in the kernel of $H'$, it should remain invariant under the projection at any two sites connected by any bond onto $\spanned\{\ket{00}+ \ket{11},\ket{0}\sigma_z(\ket{0})+\ket{1}\sigma_z(\ket{1})\}=\spanned\{\ket{00},\ket{11}\}$. 

\noindent Let us show why.

If we denote the identity by \figgs{0.15}{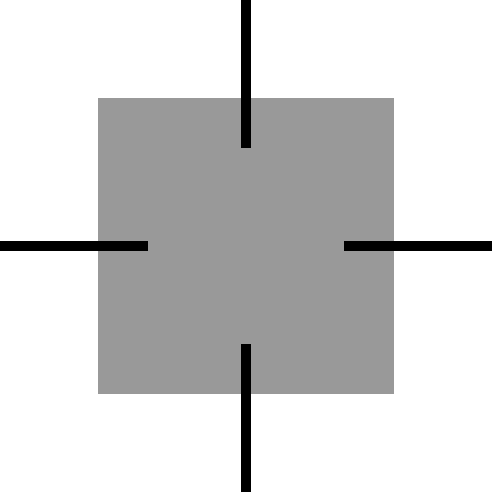}, we have
\begin{align*}
 2\  \figgs{0.15}{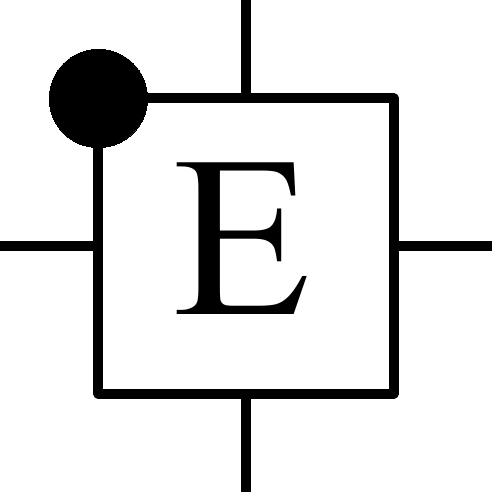}=\figgs{0.15}{pepsid}+\figgs{0.15}{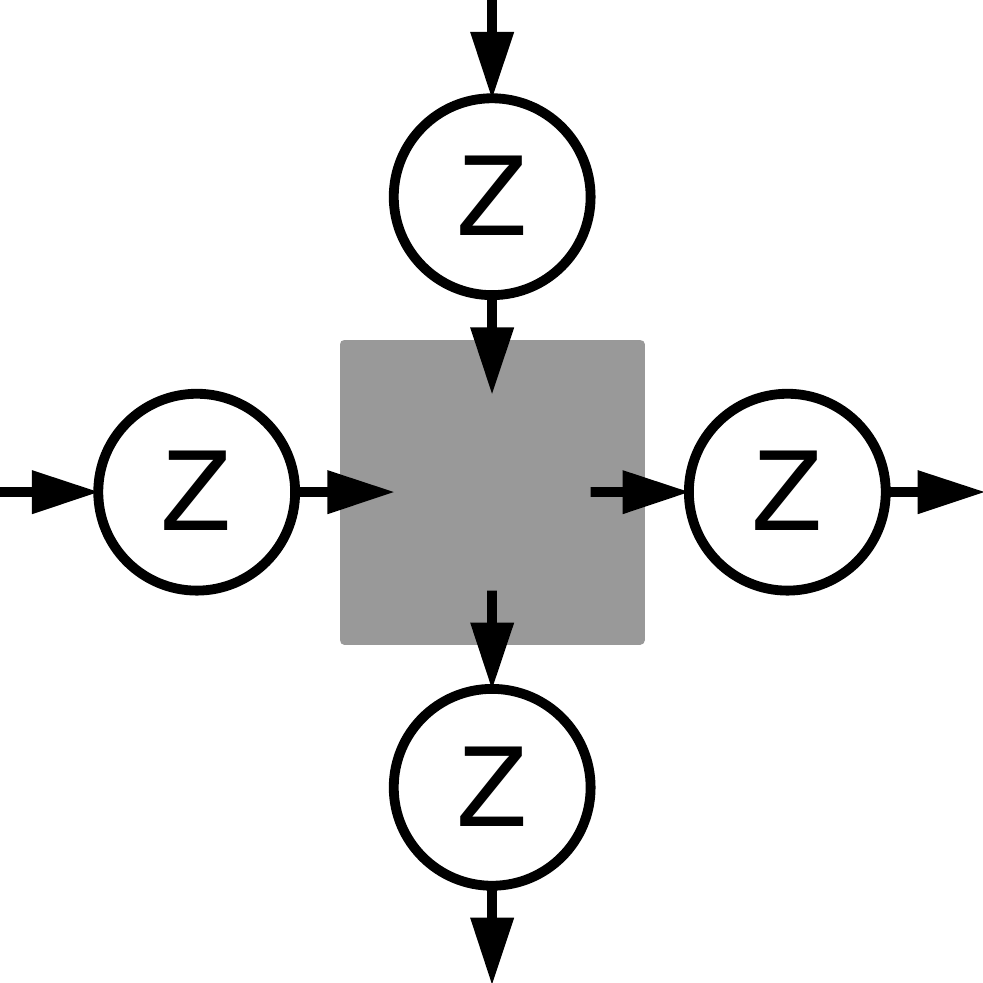} \ \ \  \hbox{ and } \ \ \ 
2\  \figgs{0.15}{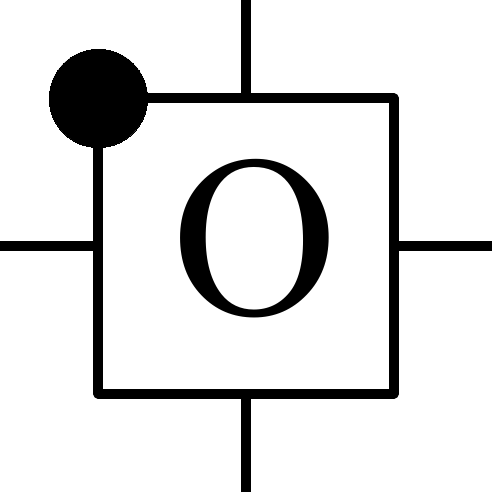}=\figgs{0.15}{pepsid}-\figgs{0.15}{pepsproj} \Rightarrow \ \ \ \  \\ 
 \Rightarrow 4 \ \figgs{0.15}{EE}\left(\hbox{alt.}\figgs{0.15}{OE}\right)=\figgs{0.15}{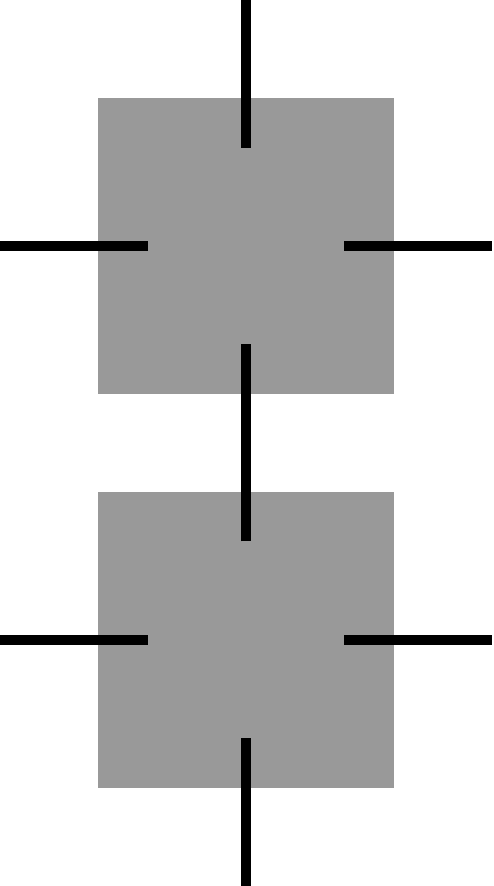}+ \figgs{0.15}{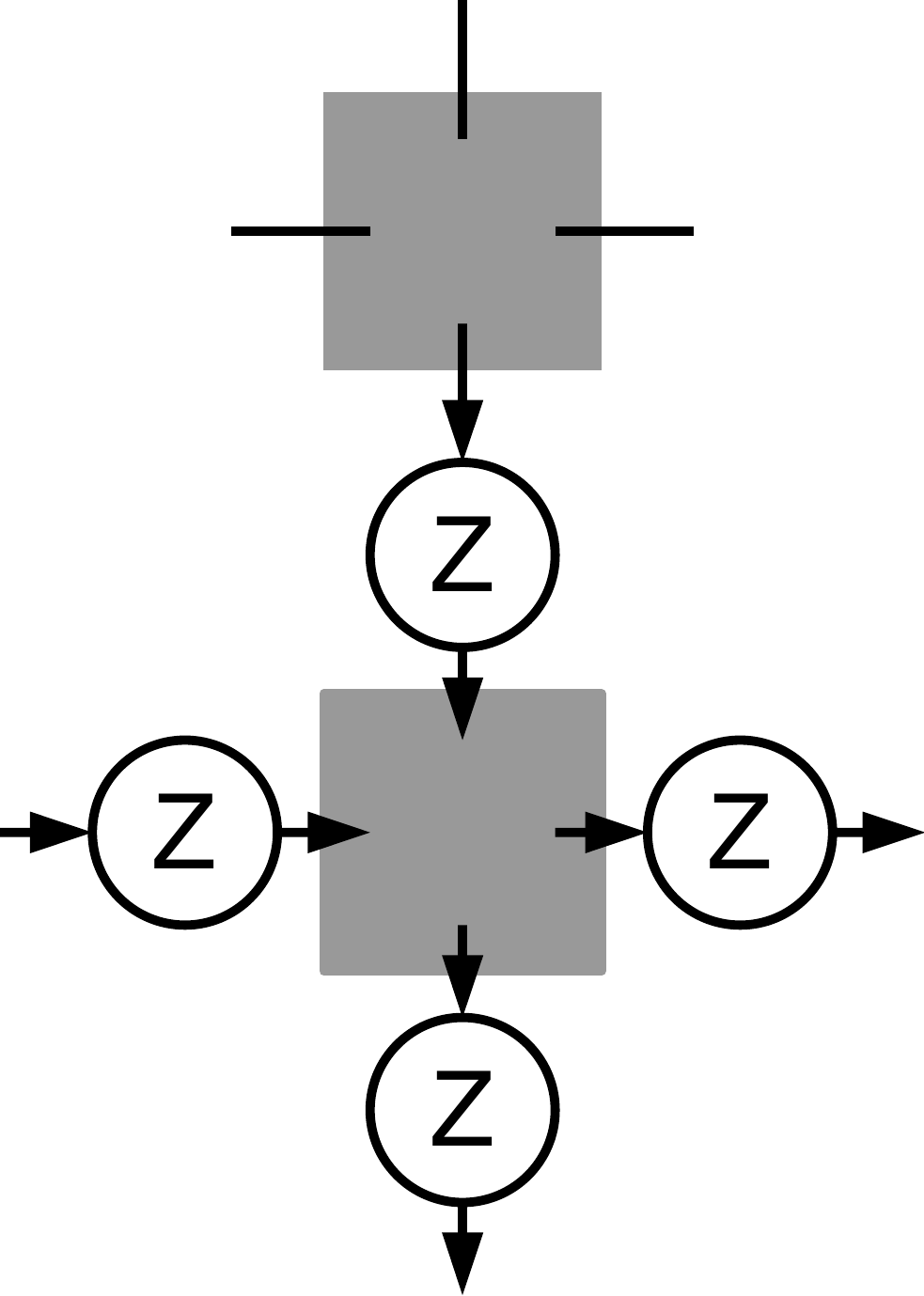}\stackrel{+ }{(-)} \figgs{0.15}{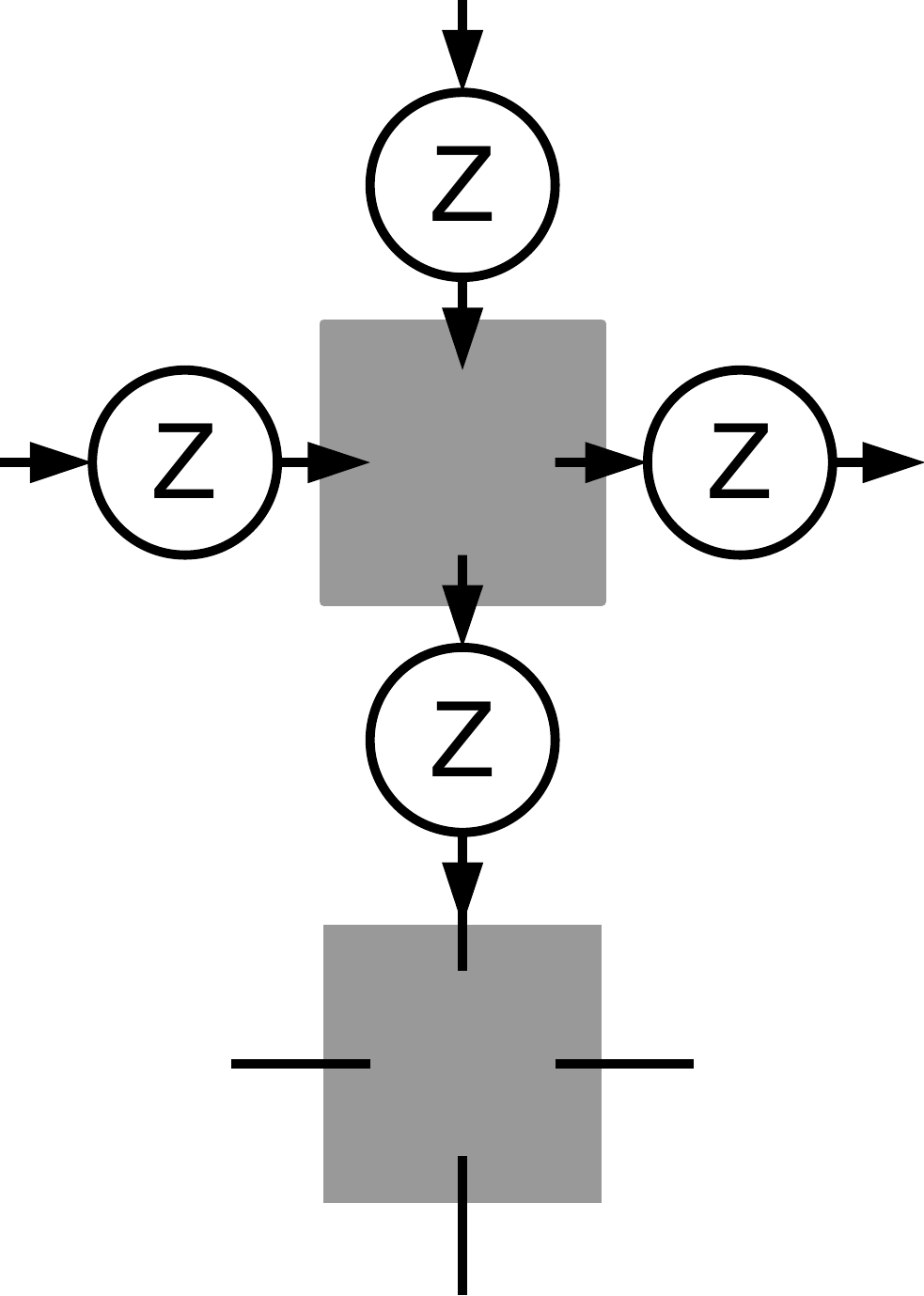} \stackrel{+}{(-)} \figgs{0.15}{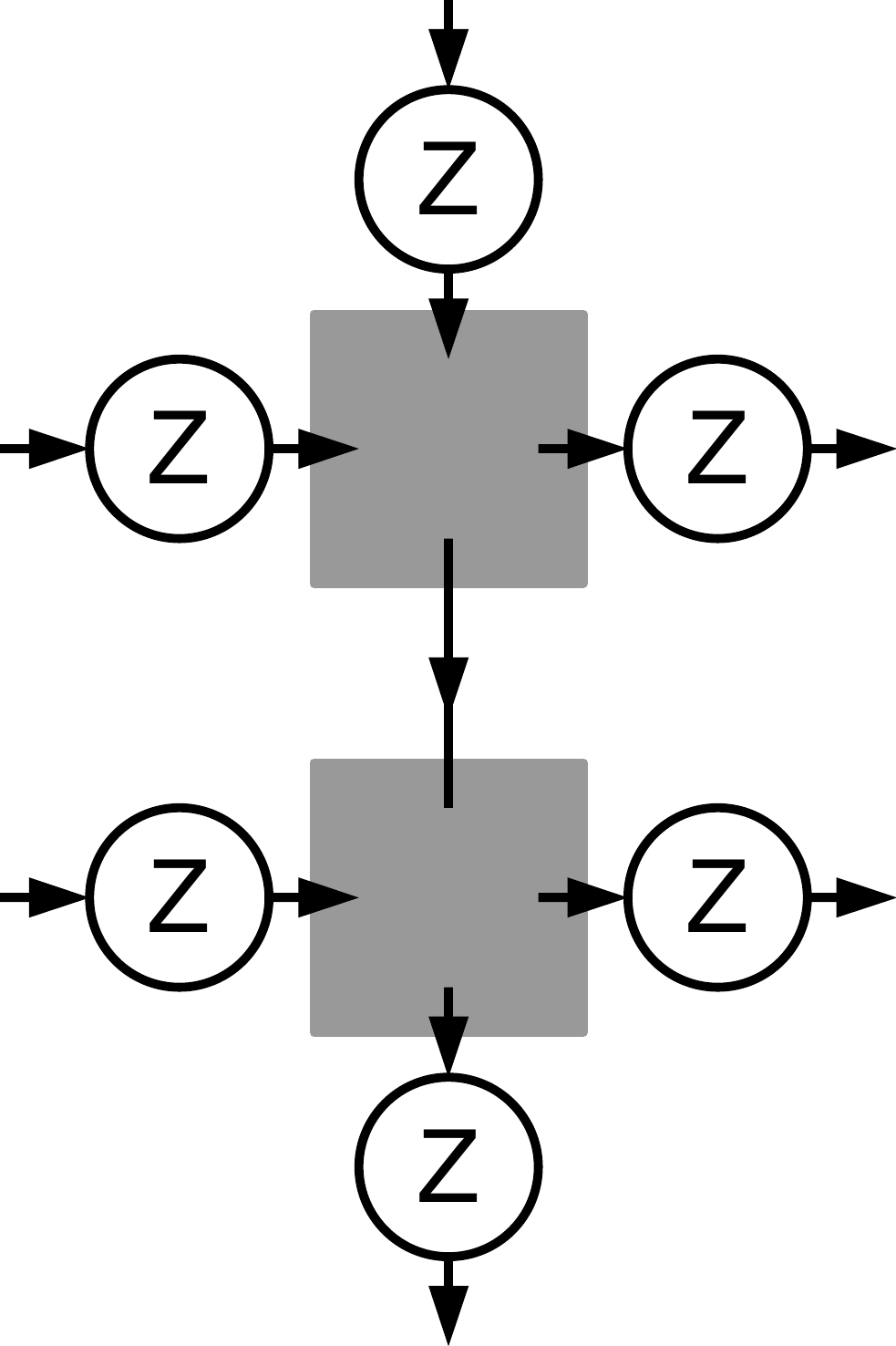}.
\end{align*}

The first and last summands remain invariant under projection onto $\spanned\{\ket{00}+\ket{11}\}$ at the sites conected by the bond, and second and third summands under projection onto $\spanned\{\ket{0}\sigma_z(\ket{0})+\ket{1}\sigma_z(\ket{1})\}$. Therefore, if we project onto the sum of these two spaces the tensors remains unchanged. 

Thus only linear combinations of the identity and $\sigma_z$ may appear in the closure bonds when imposing periodic boundary conditions, and all periodic boundary conditions are necesarily even. Hence, given the full lattice and periodic boundary conditions, the
elements in $S_{NM}$ which came from $O_{NM}$ need to vanish.

Consequently $S_{\mathrm{final}}$, the ground space of the uncle Hamiltonian, is constructed by imposing periodic boundary conditions to $E_{NM}$, and therefore coincides with the ground state subspace of the toric code parent
Hamiltonian $H_{\mathrm{TC}}$, whose detailed construction can be found in \cite{schuch:peps-sym}.  \hspace*{\fill}$\square$

\section*{Appendix B: Spectrum of the uncle Hamiltonian in the thermodynamic limit}

In this section we prove that, once we fix one of the two dimensions of the lattice, the spectrum of the uncle Hamiltonian $H'$ in the thermodynamic limit is $\R^+$. The proof follows essentially the same steps as the one from \cite{fernandez:1d-uncle} for the uncle Hamiltonian of the GHZ state. We sketch the main steps adapted to the toric code case.

The tensor $E$ appearing in this appendix is the previous one multiplied by a constant so that the MPS corresponding to a column of $E$ tensors is in its normal form \cite{perez-garcia:mps-reps}. This constant depends on the column size.

The thermodynamic limit of $H'$ can be studied as acting on the closure of the space $S=\cup _{i<j} S_{i,j}$, where
\begin{equation} \nonumber
S_{i,j}=\{\phi_{i,j}(X)=\figgs{0.15}{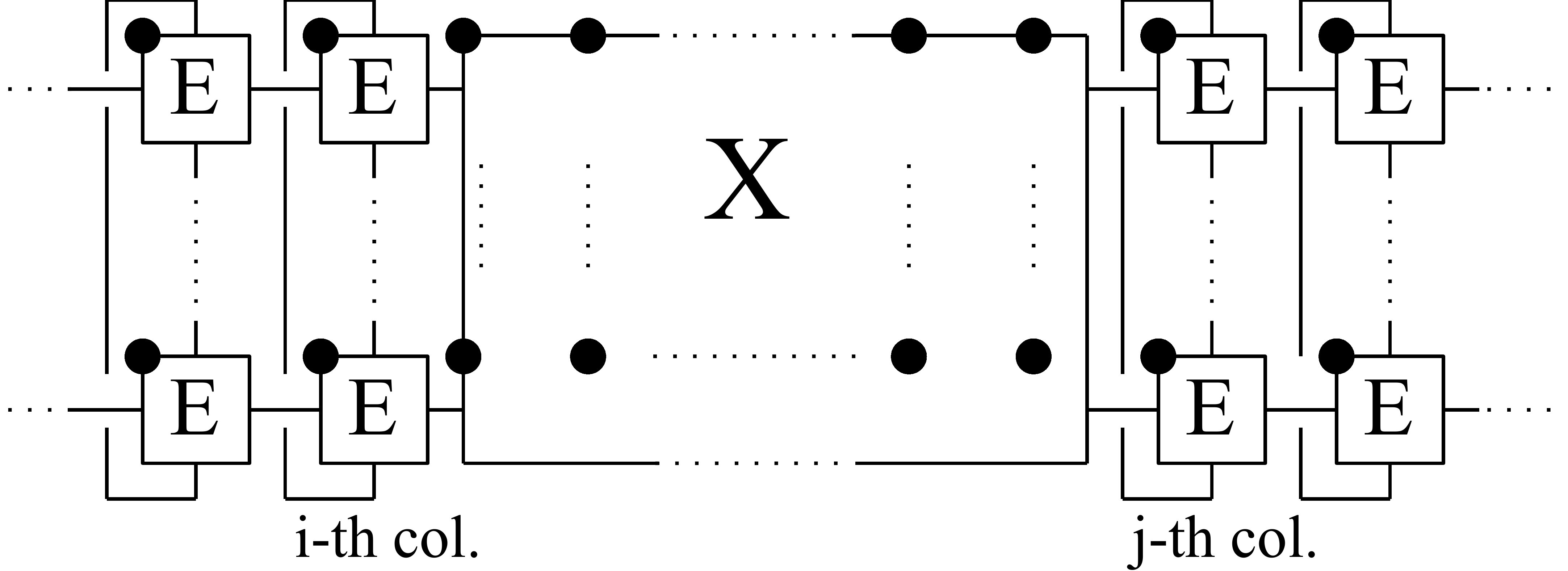},\ X\},
\end{equation}
\noindent and $X$ runs over all the possible tensors.

We will usually omit the location of $X$ whenever this does not matter due to translational invariance of the Hamiltonian.

Inside $S$ we can find the space $S^2$ spanned by vectors with the tensor $E$ everywhere but two places in which the tensor $O$ is located. In the case the tensors $O$ are located in places $(i,j)$ and $(k,l)$ of the lattice, we call this state $\ket{\phi_{i,j}^{k,l}}$.

For each of these vectors, $H'(\ket{\phi_{i,j}^{k,l}})\in \spanned \{\ket{\phi_{i+\delta_i,j+\delta_j}^{k+\delta_k,l+\delta_l}},\ \delta_i,\delta_j,\delta_k,\delta_k\in\{-1,0,1\} \}$. Therefore, $H'(S^2)\subseteq S^2$. Moreover, $H'|_{S^2}$ is bounded, and consequently it can be uniquely extended to $\overline{S^2}$, coinciding on this space with any self-adjoint extension of $H'$ to $\overline{S}$ which may exist, also called $H'$. Further study of self-adjoint extensions of unbounded symmetric operators can be found in \cite{Conway}.

{The unnormalized states $\ket{\phi_{r,N}}$, constructed as those from equation (\ref{eq:lowenergyvectors}) for rectangular $r\times N$ regions,} lie in $S^2$, and let us determine that $H'|_{\overline{S^2}}$ is gapless and there exists a sequence of elements in the spectrum $\{\lambda_i\}_i$ tending to 0. And one can find Weyl sequences in $\overline{S^2}$ associated to these values :
\be \nonumber 
\frac{\|H(\ket{\varphi_{\lambda_i,j}})-\lambda_i \ket{\varphi_{\lambda_i,j}} \|}{\|\ket{\varphi_{\lambda_i,j}}\|}
\stackrel{j\to\infty}{\longrightarrow} 0.
\ee

Using density arguments one can find these Weyl sequences lying in $S^2$. For any given $\lambda_i$ and any $\delta>0$ there exists a state $\ket{\phi_{i,\delta}}$ which is almost an eigenvector of $H'$ for the value $\lambda_i$ with an error at most $\delta$ , which means $\|(H'-\lambda_i\I)\ket{\phi_{i,\delta}}\|\le \delta \|\ket{\phi_{i,\delta}}\|$.

If we write two --or more-- of these states as $\ket{\phi_{i_1,\delta_1}}=\ket{\phi(X_1)}$ and $\ket{\phi_{i_2,\delta_2}}=\ket{\phi(X_2)}$, we can construct a new $X$ by concatenating $X_1$ and $X_2$ separated by at least two columns of $E$ tensors. We can call $\ket{\phi_2(X_1,X_2)}$ such a vector --the subindex indicates how many columns with $E$ tensors are between $X_1$ and $X_2$. This vector is an approximated eigenvector of $H'$ for $\lambda_{i_1}+\lambda_{i_2}$ with an error at most $\delta_1+\delta_2$. Let us prove that.

The first thing we need to note is that for any $\ket{\phi(X)}$ there exists a tensor $X'$ such that $H'(\ket{\phi_{i,j}(X)})=\ket{\phi_{i-1,j+1}(X')}$
\begin{equation} \nonumber
H' \left( \figgs{0.15}{EMPSb} \right)=\figgs{0.15}{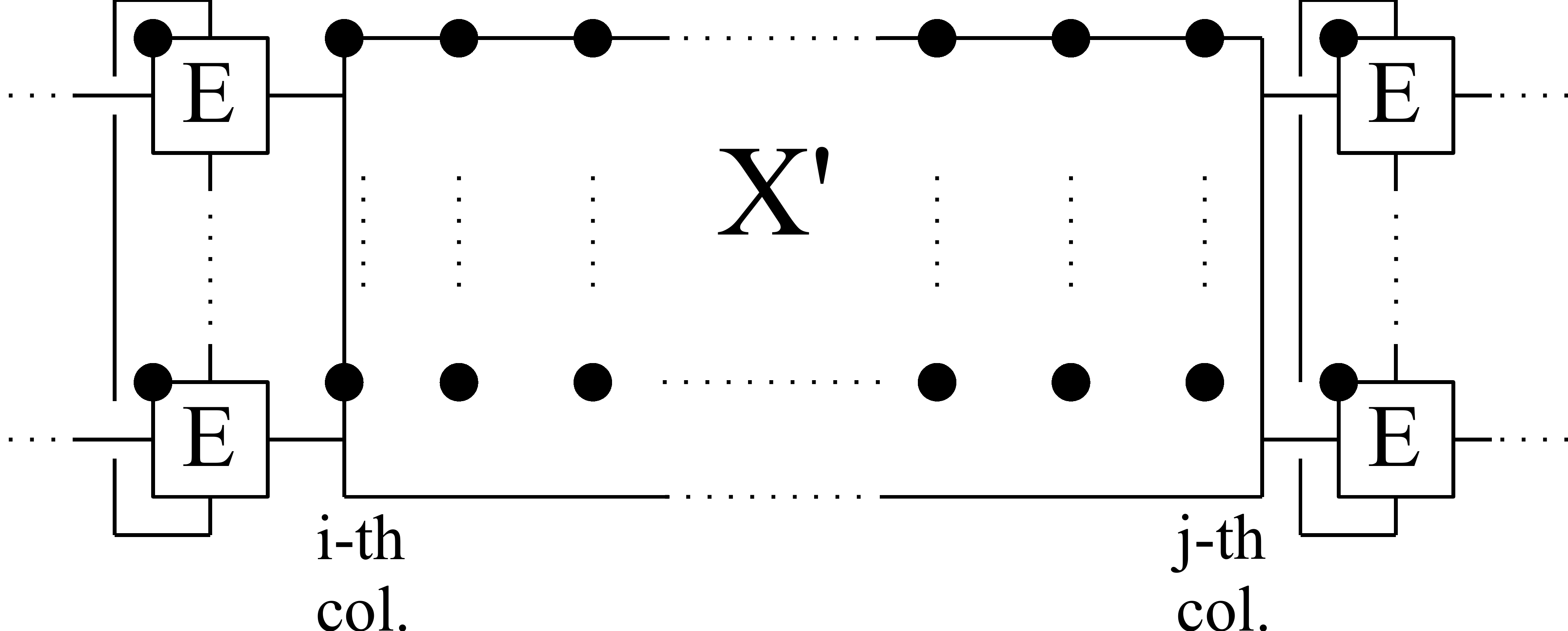}
\end{equation}

Due to the locality of $H'$, we have that $H'(\ket{\phi_2(X_1,X_2)})=\ket{\phi_1(X_1',X_2)}+\ket{\phi_1(X_1,X_2')}\stackrel{\delta_1+\delta_2}{\sim} \lambda_{i_1}\ket{\phi_2(X_1,X_2)}+\lambda_{i_2}\ket{\phi_2(X_1,X_2)}= (\lambda_{i_1}+\lambda_{i_2})\ket{\phi_2(X_1,X_2)}$.

This family of vectors let us see that any finite sum of $\lambda_i$ lies in the spectrum of $H'$. The set of finite sums of a sequence of elements tending to 0 is dense in the positive real line, and the spectrum is closed, hence $\sigma (H')=\R^+$.

The same tensors $X_i$ and $X'_i$ can be used in vectors in big enough finite dimensional lattices to show that the spectra of the uncle Hamiltonians $H'$ on finite dimensional lattices tend to be dense in the positive real line. A similar treatment is detailed in \cite{fernandez:1d-uncle}.

\end{document}